\documentclass[aps,prb,twocolumn,superscriptaddress,floatfix,english,twocolumn,amsmath,amssymmb,longbibliography]{revtex4-2}
\usepackage{graphicx}
\usepackage{dcolumn}
\usepackage{bm}
\usepackage{hyperref}
\usepackage{color}
\usepackage[T1]{fontenc}
\usepackage{amsmath}
\usepackage{amssymb}
\usepackage{esint}
\usepackage{comment}

\usepackage{amsbsy}
\usepackage{amsfonts}
\usepackage{babel}
\usepackage{enumerate}
\usepackage{epsf}
\usepackage{esint}
\usepackage{epsfig,psfrag}
\usepackage{float}
\usepackage{graphicx}
\usepackage{latexsym}

\definecolor{amaranth}{rgb}{0.9, 0.17, 0.31}
\newcommand{\buc}[1]{#1}

\begin{abstract}
	We study transport through interfaces in topological nodal-line semimetals, focusing on two \mbox{geometries}: a single interface between two large samples, one nodal-line semimetal and one metal, and an infinite nodal-line semimetal slab in between two metallic regions.
	We investigate the dependence of the spectra on the boundary conditions, showing how they affect the surface states and the band dispersion. We find a set of \emph{drum} states, arising from the hybridization of the drumhead states on opposite surfaces at finite slab width, and describe their signatures in the transport properties of a clean sample. Finally, we compute the electronic trajectories in the ballistic regime and show that there is a series of resonant angles that ensure perfect transmission. We also show how the current density profile acquires an inhomogeneous distribution in the radial direction.
\end{abstract}

\begin{document}
	
	\title{Interfaces of nodal-line semimetals: drum states, transport and refraction}
	
	\author{Mattia Rudi} 
    \email{mattia_rudi@hotmail.it}
    \affiliation{Università della Calabria, Via Pietro Bucci, 87036 Rende CS}
	\author{Alessandro De Martino}
	\email{Alessandro.De-Martino.1@city.ac.uk}
	\affiliation{Department of Mathematics, City, University of London, Northampton Square, EC1V OHB London, United Kingdom}
	\author{Kristof Moors}
    \email{k.moors@fz-juelich.de}
	\affiliation{Peter Grünberg Institute (PGI-9), Forschungszentrum Jülich, 52425 Jülich, Germany}
	\affiliation{JARA-Fundamentals of Future Information Technology, Jülich-Aachen Research Alliance, Forschungszentrum Jülich and RWTH Aachen University, Germany}
	\author{Domenico Giuliano} 	
    \email{domenico.giuliano@fis.unical.it}
    \affiliation{Università della Calabria, Via Pietro Bucci, 87036 Rende CS}	
	\author{Francesco Buccheri} 
	\email{francesco.buccheri@polito.it}
	\affiliation{Dipartimento Scienza Applicata e Tecnologia, Politecnico di Torino, Corso Duca degli Abruzzi 24, 10129 Torino, Italy}
	\affiliation{Institut f\"ur Theoretische Physik, Heinrich-Heine-Universit\"at, Universit\"atsstr. 1, D-40225  D\"usseldorf, Germany}
	
	\maketitle
	
	\section{Introduction}\label{sec:intro}
	
	The introduction of topology in condensed matter physics famously dates back several decades, yet, topological materials continue to attract renewed attention, also due to the intense experimental activity in recent years \cite{Culcer2020,Bernevig2022}. In fact, topology is nowadays an established and widely applied paradigm, promising impactful technological applications in various fields \cite{Kumar2020,Luo2023}.
	The research on topological semimetals, in particular, is fueled by their predicted and measured sizeable magnetoresistance and extremely high carrier mobility \cite{Young2012,Wang2012,Wang2013,Borisenko2014,Neupane2014,Shekhar2015}, attractive features for, e.g., ultra-sensitive detectors and fast-operating electronic devices.
	Nodal-line semimetals (NLSs) are a class of topological semimetals in which a pair of bands crosses on a one-dimensional manifold in the Brillouin zone (BZ), namely, a nodal line or nodal ring \cite{Burkov2011}. Such crossing is protected by a discrete symmetry, which quantizes the possible values of a topological invariant \cite{Fang2015,Fang2015,Chiu2014,Kim2015,Zhao2016}.
	
	The band structure of NLSs is associated to characteristic signatures in a number of phenomena \cite{Hu2017,Rui2018,Yang2022}, including quantum oscillations from the toroidal shape of the Fermi surface \cite{Yang2018,Li2018,Oroszlany2018}, the anomalous Hall effect in magnetic materials \cite{Kim2018}, ultra-flat bands in magnetic field and related magnetotransport \cite{Rhim2015}.
	There are, to date, numerous proposed materials to host a NLSs phase and mounting experimental confirmation  \cite{Xie2015,Neupane2016,Bian2016p,Schoop2016,Hu2016,Hosen2017,Chen2017,Takane2018,Liu2018x,Qiu2019,Fu2019,Belopolski2019,Song2020,Muechler2020,Wang2021,Zhang2022nls}. Moreover, it has been shown that phononic and photonic crystals can be engineered to form synthetic analogues of nodal-line materials \cite{Deng2019,Park2022}, and can be useful to probe all those characteristics that only depend on the eigenenergies. 
	It was observed, in particular, that the carrier mobility can reach sizeable values in clean samples, orders of magnitude higher than typical metals at room temperature and comparable to that of graphene \cite{Laha2020}.
	Interestingly, the relative flatness of the band along the nodal ring makes NLSs also a good platform for highlighting correlation effects \cite{Shao2020,Heikkila2016,Nandkishore2016} and superconductivity \cite{Pereira2019,Yamada2021,Shang2022}. Strongly-correlated nodal-line semimetals have recently been synthesized in the laboratory \cite{Chen2022}. 
	
	Topological surface states are generally expected when two materials with a different value of a bulk topological invariant are in contact \cite{Hasan2010}. In NLSs, they have support on the disk delimited by the projection of the nodal line onto the interface BZ: for this reason they are dubbed "drumhead" states in the literature \cite{Burkov2011,Chan2016a}. Such states are present even at non-ideal interfaces, and are associated with van Hove singularities in the density of states \cite{Chan2016a,Buccheri2024}, robust signatures in spectroscopy \cite{Bian2016d,Fu2019,Belopolski2019,Muechler2020}, quasiparticle interference \cite{Biderang2018,Stuart2022} and spin-polarized transport \cite{Chen2018}.
	A small dispersion of surface bands can arise from the particle-hole symmetry breaking, manifest in the dispersion of the nodal line itself \cite{Burkov2011,Weng2015,Yu2015,Rui2018}.
	
	In this work, we consider interfaces between a NLS and the vacuum or another material, which are always parallel to the plane of the nodal line. 
	We describe the most general parametric family of boundary conditions preserving self-adjointness and mirror symmetry, in which the parameter is related to the surface composition. We show that the boundary parameter determines the penetration length and the dispersion of the surface band: as our model does not break particle-hole symmetry in the bulk, we conclude that the surface states can acquire a dispersion even in the absence of a tilt of the nodal line.
	
	We also show that, in a slab geometry, the drumhead states on opposite surfaces can hybridize, forming a pair of "drum" states. \buc{Despite their exponential localization, such states can produce a sizeable contribution in transport measurements across the slab,
  because their penetration length becomes comparable to the sample size in the region around the nodal line.}
  
	Finally, we also explore the consequences of the electron refraction, originated by the change of dispersion at the interface. The possibility of electron focusing in a crystalline solid was experimentally demonstrated in ballistic two-dimensional electron gases at semiconductor interfaces \cite{Spector1990,Sivan1990}, paving the way for all subsequent electron optics studies. Most famously, the Veselago lensing in graphene has been observed \cite{Cheianov2007,Chen2016,Lee2015,Taychatanapat2013}, taking advantage of the very long mean free path in the material. 
	In a sufficiently clean topological semimetal sample, measured mean free paths can be of the order of tens of $\mu m$ \cite{Perevalova2022}. For this reason, they offer very good candidates for observing focusing effects in three dimensions \cite{Buccheri2022b} and the possibility was indeed already demonstrated in a photonic lattice Weyl semimetal \cite{Yang2021}.
	In nodal-line semimetals, specifically, electronic correlations play a comparatively larger role than in other topological semimetals \cite{Shao2020}, reducing the characteristic scattering time and renormalizing the electron velocity. While at low carrier density and temperature the Coulomb interaction has a power law behavior at short distances, at larger distances
	or at sufficiently high temperature or chemical potential, the interaction is exponentially suppressed \cite{Syzranov2017} and an analysis based on a single-particle Hamiltonian is approximately justified.
	In this work, we adopt such a viewpoint and show that, in a ballistic junction, the band dispersion associated with a nodal line refracts part of an incident wavepacket toward the main axis. We also propose an experiment that exploits the electron optical properties of the NLS to image the nodal line in real space.
	
	The paper is structured as follows. In Sec. \ref{sec:NLS}, we introduce our model and its eigenstates, then proceed to imposing the most general boundary conditions and describe the dispersion and penetration length of the surface states, as well as their impact on the transport properties of the surface. In Sec. \ref{sec:N-NLS}, we study the electronic tunneling through a single interface between a metallic material and a NLS and we proceed to study the transport through a NLS slab in Sec. \ref{sec:N-NLS-N}, with particular focus on the resonances. We also tackle the current distribution and demonstrate the possibility of focusing the transmitted electrons on a ring, as well as of increasing an electron beam collimation over a certain region by using a NLS slab. We conclude by summing up our considerations in Sec. \ref{sec:conclusions}. A few complementary technical details are provided in Appendices.

	\section{Model of a nodal-line semimetal: bulk and interfaces}\label{sec:NLS}
\begin{figure}
		\includegraphics[height=0.2\textheight]{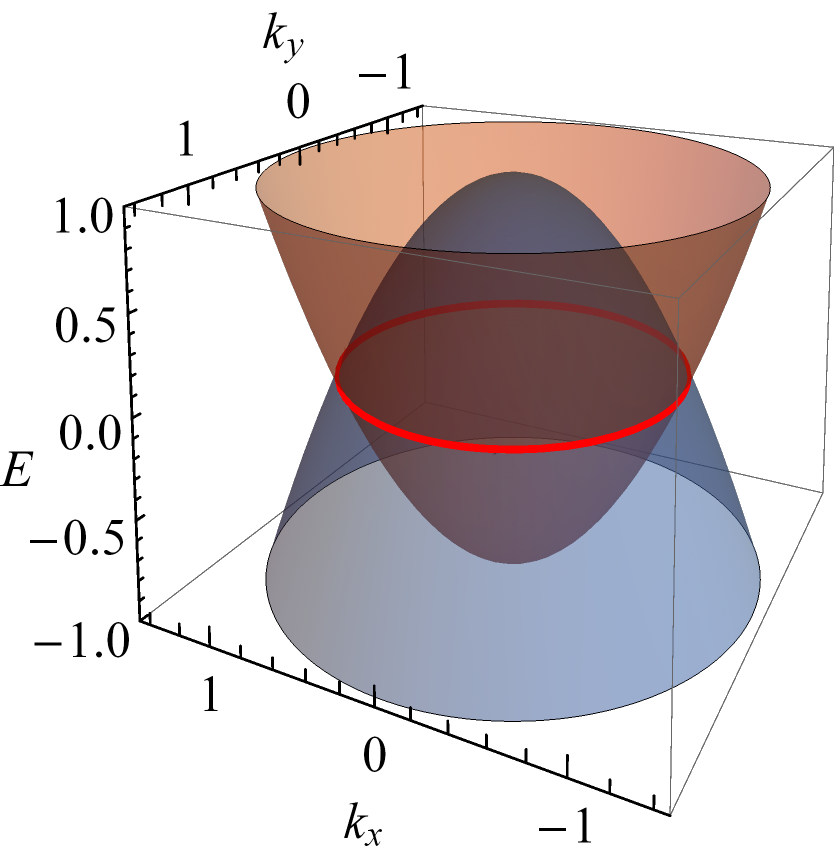}
		\caption{
			\buc{Schematic representation of the energy bands \eqref{eq:nrg} (in arbitrary units) in the NLS semimetal regime: a ring-shaped crossing with radius $k_\mathrm{NL}$ (here with unit value) is present at zero energy.} \label{fig:Bands} 
		}
	\end{figure}

	We start by considering an effective Hamiltonian \buc{describing the vicinity of the nodal line} of the form \cite{Burkov2011,Chan2016a}
	\begin{equation}
		H  =  
		-i \hbar v \partial_{z}\tau_{y} +  \mathcal{M}\left(k_\mathrm{p}\right)\tau_{z}+\mathcal{V},
		\label{eq:H}
	\end{equation}
	where $\tau_{j}$, $j=x,y,z$ denote the Pauli matrices, acting on an effective degree of freedom that can be represented on a spinor. In the example of $\mbox{Ca}_3\mbox{P}_2$ \cite{Xie2015}, the two bands appearing in  Eq. \eqref{eq:H} are formed out of the $p$ and $d$ orbitals of $P$ and $\mbox{Ca}$, respectively. \buc{The scalar term $\mathcal{V}$ describes a global shift of the bands. In Section \ref{sec:N-NLS} we will use a space-dependent $\mathcal{V}(z)$  to model an interface between a NLS and a metal.} The Hamiltonian is fully invariant under $SU(2)$, hence, the spin degree of freedom does not appear explicitly. The momentum components parallel and perpendicular to the plane identified by the nodal-line are denoted as $\boldsymbol{k}_\mathrm{p}\equiv\left(k_x,k_y\right) $ and $k_z$, respectively.
	The function 
	\begin{equation}\label{eq:m}
		\mathcal{M}\left(k_\mathrm{p}\right)\equiv\mathcal{D}\left(k_\mathrm{p}^{2}-a\right)
	\end{equation}
	of  $k_\mathrm{p}\equiv|\boldsymbol{k}_\mathrm{p}|$ has a zero at $k_\mathrm{NL}\equiv\sqrt{a}$, provided \mbox{$a>0$}.  More general shapes of the nodal ring can be modeled by introducing an angular dependence of $a$, which is tackled via a reparametrization of the coordinates and momenta in the $x$$y$ plane. The eigenvalues of the Hamiltonian \eqref{eq:H} \buc{for constant $\mathcal{V}$} are
	\begin{equation}\label{eq:nrg}
		E_\pm\left({\boldsymbol{k}} \right) = \pm\sqrt{\hbar^2 v^2 k_z^\buc{2}+\mathcal{M}^2\left(k_\mathrm{p}\right)}+\mathcal{V}\, .
	\end{equation}
    \buc{and are shown in Fig. \ref{fig:Bands} for sample values of the parameters.} 
	When $\mathcal{V}=0$ and $a>0$, this Hamiltonian describes a NLS with a nodal line at $k_z=0$, $k_\mathrm{p}=k_\mathrm{NL}$, where the two bands cross at zero energy. The energy $2E_0\equiv2\mathcal{D}|a|$ corresponds to the maximal separation between the band in the inverted gap region if $a>0$ and it is typically a fraction \footnote{For instance, $E_0\approx0.2\mbox{ eV}$ in $\mbox{Ca}_3\mbox{P}_2$} of $\mbox{eV}$ \cite{Chen2016,Takane2018,Qiu2019}. This is a natural ultraviolet cut-off energy scale for our effective model and is a material-dependent parameter. 

	The Hamiltonian in Eq. \eqref{eq:H} can also model an insulator with band gap $\Delta=2E_0$ whenever $a<0$. Moreover, it has two metallic regimes (N), which can be reached by shifting the bands with the term $\mathcal{V}$. At zero chemical potential, for $\mathcal{V}\ll -E_0$, it features electron-like carriers and a spherical Fermi surface, while for $\mathcal{V}\gg E_0$, one has a metal with hole-like carriers. As seen in Sec. \ref{sec:N-NLS}, the geometry of the Fermi surface is of importance in describing the transmission properties of a N-NLS junction.
	
	The eigenstates corresponding to Eq. \eqref{eq:nrg} can be written in the form
	\begin{eqnarray}\label{eq:eigenstates}
		\psi_{\boldsymbol{k}}=\mathcal{N}_{\boldsymbol{k}}
		\left(\begin{array}{c}
			-i \hbar v k_{z}\\
			E-\mathcal{V}-\mathcal{M}\left( k_\mathrm{p} \right) 
		\end{array}\right)
	\end{eqnarray}
	with $\mathcal{N}_{\boldsymbol{k}}^{-1}\equiv\sqrt{2\left(E-\mathcal{V} \right)\left(E-\mathcal{V}-\mathcal{M}\left( k_\mathrm{p} \right) \right)}$ the normalization factor.
	The Hamiltonian \eqref{eq:H} is not a good description for the whole BZ, but only in the vicinity of the nodal line. As it will not affect our conclusions, we will assume throughout this work that the description holds, at least qualitatively, also for $\boldsymbol{k}\to0$, while this is not the case close to the edge of the BZ. We will comment on the specific limitations as they arise.
	
	Importantly, the Hamiltonian \eqref{eq:H} possesses, together with particle-hole \mbox{$H\left( \boldsymbol{k} \right)=-\tau_x H \left( -\boldsymbol{k} \right)^* \tau_x$} \buc{(for $\mathcal{V}=0$)} and time-reversal \mbox{$H\left( -\boldsymbol{k} \right)=
 H\left( \boldsymbol{k} \right)^*$} symmetries, a mirror symmetry with respect to the plane $z=0$, implemented as 
    \begin{equation}\label{eq:mirror}
        H\left(k_z\right)=\tau_z H\left(-k_z\right)\tau_z \;.
    \end{equation}   
	The presence of this discrete symmetry squaring to the identity protects the nodal line and the corresponding surface states \cite{Zhao2016,Chan2016a,Rui2018}.
	
	Without loss of generality, we will measure energies with respect to the energy of the band crossing in the NLS and, in order to lighten the notation, we shall work with dimensionless quantities in the following. In our convention, we express all energies ($\varepsilon$) in units of $\hbar v k_\mathrm{NL}$ and momenta ($\mathbf{q}$) in units of $k_\mathrm{NL}$, i.e., $q_\mathrm{p}=1$ will identify the nodal line. Analogously, we define the dimensionless "mass" function $m\left(q_\mathrm{p}\right)\equiv \mathcal{M}\left(k_\mathrm{p}\right)/\hbar v k_\mathrm{NL}=\buc{D\left(q_p^2-1\right)}$, where $D\equiv k_\mathrm{NL}\mathcal{D}/\hbar v$. For later use, we introduce the notation $V\equiv \mathcal{V}/\hbar v k_\mathrm{NL}$, as well as the sample width $\mathcal{L}$ and its dimensionless version $L \equiv k_\mathrm{NL}\mathcal{L}$.

    \buc{To conclude this discussion we note that, while for simplicity we use a particle-hole symmetric Hamiltonian, a term which breaks this symmetry can in general be present. In the case of a scalar function of the momenta \cite{Chan2016a}, it can be readily taken into account by means of a suitable  transformation and a rescaling of the eigenenergies \cite{Buccheri2024}. Therefore, we do not expect a qualitative changes to our results.}

	\subsection{Interface with the vacuum}\label{sec:obc}
	In order to gain insight on the role of boundaries in topological semimetals, we now consider a 
    surface at $z=0$. The results of this section will form the basis for the conductance calculations in Sec. \ref{sec:N-NLS}.
	For an infinitely-extended system in the $z$ direction, we expect two-dimensional drumhead surface bands, with flat dispersion at zero energy. These states have support within a circle in momentum space, delimited by the projection of the nodal line onto the surface BZ, and penetration length that diverges as $1/\left(1-q_\mathrm{p}\right)$ when the in-plane momentum approaches the nodal ring \cite{Burkov2011,Jackiw1976}.
	Because of this, it is reasonable to expect that finite-size effects will play an important role. Similarly to Weyl semimetals, the most general boundary conditions for a Dirac Hamiltonian that can be imposed on the surface $z=0$ are of the form 
	\begin{equation}\label{eq:bc}
		B\left(\alpha\right) \psi_\alpha\left(0\right) = \psi_\alpha\left(0\right)\;,
	\end{equation}
	where
	$B\left(\alpha\right) = \tau^z \cos\alpha+\tau^x \sin\alpha$
	is a one-parameter (the angle $\alpha$) family of Hermitian matrices, that ensures the self-adjointness of the Hamiltonian \eqref{eq:H}  \cite{Witten2016}. Using the fact that $B\left(\alpha\right)$ anticommutes with the current $j_z=ev\tau_y$, it can be shown that the expectation value of $j_z$ vanishes on the surface at $z=0$: for this reason, these conditions are sometimes referred to as zero-current boundary conditions.
	As underlined in Ref. \cite{Chan2016a}, the Dirac Hamiltonian with the linear term $k_z$ only establishes that the gapped regions inside and outside the nodal line are topologically distinct. The nodal line separates two regions of the BZ with a different value of the bulk topological invariant, defined by the Zak's phase along $q_z$ for fixed $\mathbf{q}_\mathrm{p}$. In order to resolve which of these regions supports boundary states, higher-order terms are, in general, necessary \cite{Shen2017}. Nevertheless, as we show in this work, the boundary conditions determine where the topological surface states are present and, therefore, define the topologically nontrivial region. With physical systems in mind, throughout this work, we will describe the surface states inside the nodal line, with the understanding that the alternative range of the boundary parameter leads to the other configuration of the surface states.
	
	The angle $\alpha$ in Eq. \eqref{eq:bc} has been introduced as a way of guaranteeing the formal consistency of the low-energy Hamiltonian. It models aspects of the surface that are not explicitly taken into account in the low-energy Hamiltonian. In the context of Weyl semimetals, for instance, it accounts for the curvature of the Fermi arcs \cite{Buccheri2022a}. Provided the surface can be considered homogeneous on a macroscopic scale and invariant under translations, aspects like, e.g., the chemical composition of the termination and the electrostatic fields on the surface \cite{Morali2019,Yang2019,Ekahana2020} can be captured by the single phenomenological parameter $\alpha$.
	In the fully $SU(2)$-invariant model of $\mbox{Ca}_3\mbox{P}_2$ \cite{Xie2015,Chan2016a}, the interpretation of the two bands in Eq. \eqref{eq:H} is that of the $p-$ and $d-$ orbitals of the two components of the binary compound. Then $\alpha=0$ corresponds to a "polarization" in orbital space along the \mbox{Ca} $d$-type orbitals, i.e., a surface termination by that element. Analogously, the termination with \mbox{P} on the surface ($p$-type orbitals) is modeled by $\alpha=\pi$, while a generic value of $\alpha$ denotes a termination that contains both elements.
	One way of connecting this phenomenological parameter to experiments that can determine it in a material-independent way is proposed in Sec. \ref{sec:surfacetransport}.
	
	We now derive the dispersion of the surface states in the presence of general boundary conditions. The eigenstates \eqref{eq:eigenstates}, indeed, can be shown to describe localized states as well, under analytic continuation $k_z=i\kappa_z$: the constraint \eqref{eq:bc} fixes the spinor orientation on the surface as $\psi \propto \xi_{\alpha+}$, having denoted with 
	$\xi_{\alpha,\pm}$ the eigenstate of $B\left(\alpha\right)$ with eigenvalue $\pm1$ (see also App. \ref{app:quantization}).
	For $V=0$ and $0<\alpha<\pi$, one finds a two-dimensional band corresponding to localized states of the form
	\begin{equation}\label{eq:drumheadstate}
		\psi_{\mathbf{q}_\mathrm{p}}\left( \boldsymbol{r}\right) 	= \sqrt{\frac{\kappa}{2\pi^2}} \left( \begin{array}{c}
			\cos\left( \alpha/2\right) \\
			\sin\left( \alpha/2\right)
		\end{array} \right) e^{-\kappa  z+iq_xx+iq_y y}
	\end{equation}
	with support in the interval $0\le q_\mathrm{p}^2<1$, i.e., a disk in the surface BZ.
	Here the inverse penetration depth is given by
	\begin{equation}\label{eq:kappainf}
		\kappa=\kappa\left(q_\mathrm{p}\right)=-m\left(q_\mathrm{p}\right) \sin\alpha
	\end{equation}  and the corresponding dispersion relation reads
	\begin{equation}\label{eq:ssdispersion}
		\varepsilon\left( {q}_\mathrm{p}\right) = m\left(q_\mathrm{p}\right)\cos\alpha \;,
	\end{equation}
	and is not flat for generic values of $\alpha$. These states describe drumhead states, filling the projection of the nodal line onto the surface BZ. For the choice $\alpha={\pi}/{2}$, they reproduce the known states in the flat surface band \cite{Burkov2011}.
	
	\subsection{Boundary conditions and surface transport}\label{sec:surfacetransport}
	\begin{figure}
		\includegraphics[height=0.1\textheight]{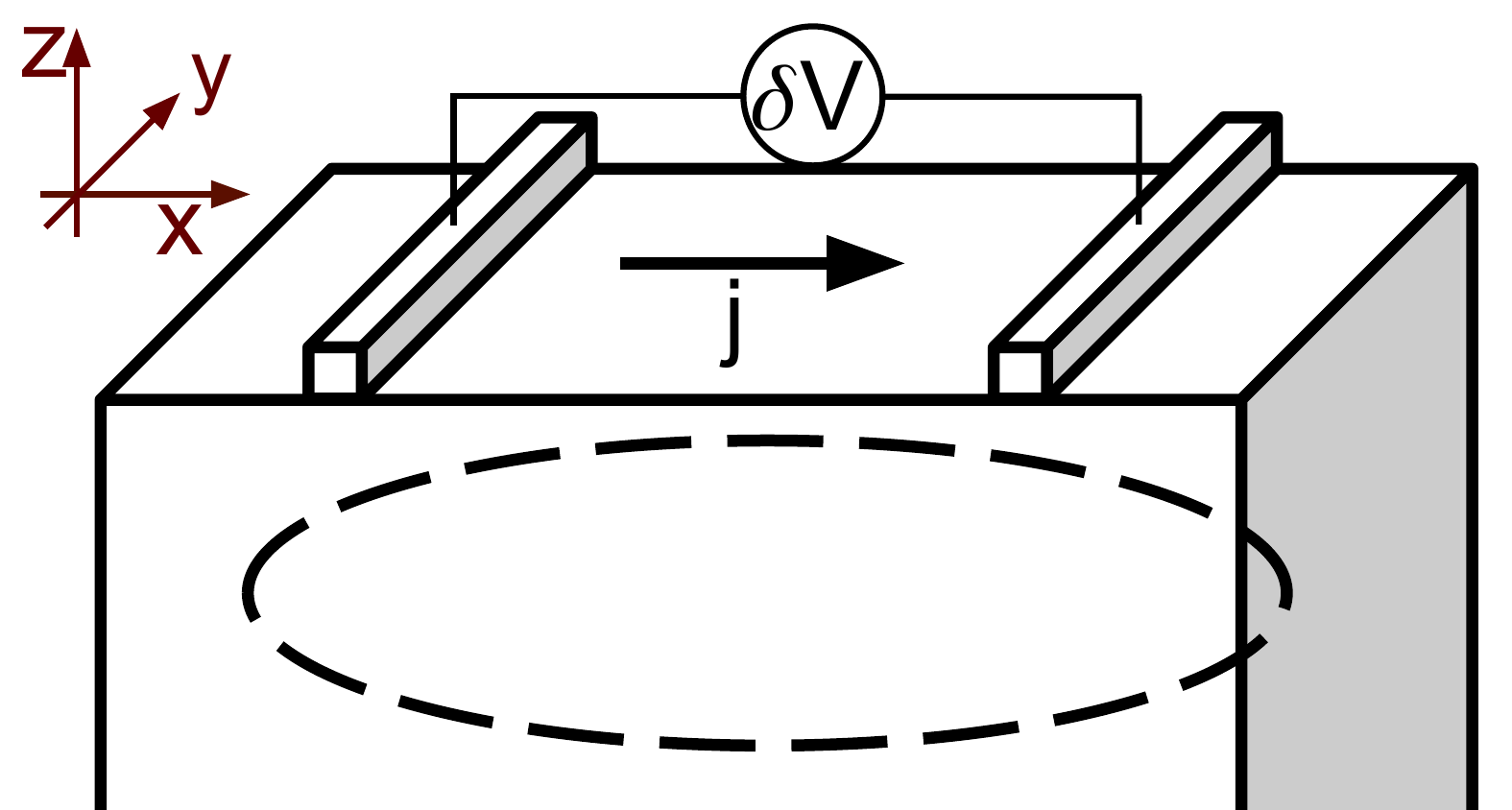}
		\caption{
			Schematic of the proposed surface measurement, with two infinitely-extended contacts acting as source and drain on the surface of a NLS, parallel to the plane that hosts the nodal line. The current in linear response depends on the boundary parameter $\alpha$ as discussed in the main text. \label{fig:surfacegates} 
		}
	\end{figure}
	We now examine the effects of the boundary parameter. In order to do this, consider the setup in Fig. \ref{fig:surfacegates}, in which a pair of contacts probes the current on the surface of the sample hosting the topological state. We consider a NLS slab, with area $\mathcal{S}$ in the $xy$ plane and with size $\mathcal{L}$ along $z$, which is for the the moment used as a regulator and considered to be large. For simplicity, we will limit our analysis to the regime $0<\alpha\le \pi/2$.
	The bulk density of states  \buc{for the Hamiltonian \eqref{eq:H}} \mbox{ $n^{(\text{b})}=\frac{1}{\mathcal{L}\mathcal{S}}\sum_{\boldsymbol{k},\buc{\pm}}\delta\left(E-E_\pm\left(\boldsymbol{k}\right) \right)=\frac{\left|E\right|}{2\pi\hbar v \mathcal{D}}$} vanishes linearly at the nodal line, hence, in the ideal case in which the chemical potential is aligned with the band crossing, the only contributions must come from the surface. Presently available samples of nodal-line semimetals have non-vanishing bulk carrier densities, hugely varying between $10^{16}-10^{20}\mbox{cm}^{-3}$ \cite{Hu2016,Hu2017,Emmanouilidou2017,Laha2019,Laha2020,Wang2022a}, still orders of magnitude lower than metals. It is still useful to consider the situation in which $n^{(\text{b})}\approx 0$, while the surface component of the density of states is
	\begin{equation}\label{eq:surfaceDoS}
		n^{(\text{s})}(E)
		=\frac{1}{4\pi\mathcal{L}\mathcal{D}\cos\alpha}\Theta\left( a\mathcal{D}\cos\alpha+E\right) \Theta\left(-E\right) \,.
	\end{equation}
    where $\Theta$ denotes the Heaviside step function. For large widths and generic values of $\alpha$, the bulk states give the dominant contribution to the density of states. On the other hand, the two contributions become comparable as soon as \mbox{$\left|E\right|\mathcal{L}\cos \alpha  \lesssim 2\pi \hbar v$}, and the surface density of states eventually becomes dominant when the inequality is strictly satisfied. In the flat-band limit, one obtains a divergent DOS in the limit $E\to0$ \cite{Burkov2011}, \buc{which has been argued to generate an enhanced Josephson current through a NLS sample \cite{Parhizgar2020}}.

	The boundary angle $\alpha$ determines the degree of penetration of the drumhead states in the sample, as seen from the decay length in Eq. \eqref{eq:kappainf} and the local density of states (per unit surface)
 \begin{align}\label{eq:surfaceLDoS}
     A_s\left(z;E \right) =\;&
     \frac{|E|\sin\alpha e^{2 z E \tan\alpha/\hbar v}}{2\pi \hbar v k_\mathrm{NL}\mathcal{D}\cos^2\alpha}
     \\
     &\;\times
     \Theta\left(-E\right)\Theta\left(\mathcal{D}a\cos\alpha+E\right) .\nonumber
 \end{align}
	We now show that the angle $\alpha$ has observable consequences in transport experiments, and start by noting that the semiclassical velocity of the surface electrons is \mbox{$\boldsymbol{v}=\boldsymbol{v}\left(\boldsymbol{q}_\mathrm{p}\right)=2D\boldsymbol{q}_\mathrm{p}\cos\alpha$}, thus, contains explicitly the boundary parameter. 
	
	Let us assume that the carriers in the electrodes are in equilibrium at the same temperature $T$, as well as a slight imbalance in chemical potential, manipulated through a voltage bias, that we can treat in linear response. We apply a standard Boltzmann formalism in the relaxation-time approximation, under the hypothesis of a weak relaxation rate due to the residual interactions and scattering with impurities or phonons.
	With the electric field polarization $\hat{u}$, the surface conductivity in linear response is
	\begin{equation}\label{eq:sigmas}
		\sigma_s=e^2\intop \frac{d^2\boldsymbol{k}_p}{\left( 2\pi\right)^2 } \tau \left(\boldsymbol{v}\cdot\hat{u} \right)^2\left(-\frac{\partial f}{\partial E} \right) = \frac{e^{2}}{h}\frac{\mathcal{D}\tau_0k_\mathrm{NL}^2}{\hbar}\cos\alpha\;.
	\end{equation}
	In this expression, we have approximated the relaxation time with its value at the Fermi energy $\tau_0$, treated here as a phenomenological parameter, obtaining a temperature-independent result for $k_BT\ll \hbar v k_\mathrm{NL}$. The latter can be estimated as a fraction of \mbox{eV} (in the example of $\mbox{Ca}_3\mbox{P}_2$, $\hbar v k_\mathrm{NL} = 0.515$ \mbox{eV} \cite{Xie2015}). \buc{The scattering time $\tau_0$ is determined by a complex interplay of electron-electron interactions, electron-phonon interactions, and impurity scattering, involving both surface as well as bulk states. 
    Here, we treat it as a phenomenological parameter, independent of temperature and surface termination.}
	
	While this is a crude estimate, which can be affected by fluctuations of the bulk chemical potential and by surface-bulk coupling \cite{Buccheri2022a}, we can nonetheless draw some interesting conclusions from our analysis. First of all, we note that the same system, even in the presence of a certain degree of protection of the surface states due to the bulk symmetries, can show very different responses in the surface conductivity, depending on the details of the surface itself. In systems for which the bands in Eq. \eqref{eq:H} represent orbital types \cite{Chan2016a}, the expression above links the chemical composition of the surface to its charge conductivity. Specifically, if the Hamiltonian \eqref{eq:H} is written in the basis of the elements of a binary compound, the relative concentration on the surface is $\tan\left(\alpha/2\right)$. Then, the surface conductivity is proportional to $\cos\alpha$, which, in turn, implies that surfaces with one element type only are relatively bad conductors. 
	The (electronic contribution to the) low-temperature thermal conductivity $k_s$ is related to the electric conductivity by the Wiedemann-Franz law \cite{Grosso}, hence, $k_s=\frac{k_B^2T}{e^2}\sigma_s + O\left(T^3\right)$ exhibits the same dependence from the boundary angle.

	\section{N-NLS interface}\label{sec:N-NLS}
	
	\subsection{Model of the interface}\label{subsec:intmodel}
\begin{figure}
		\centering
		\includegraphics[height=0.17\textheight]{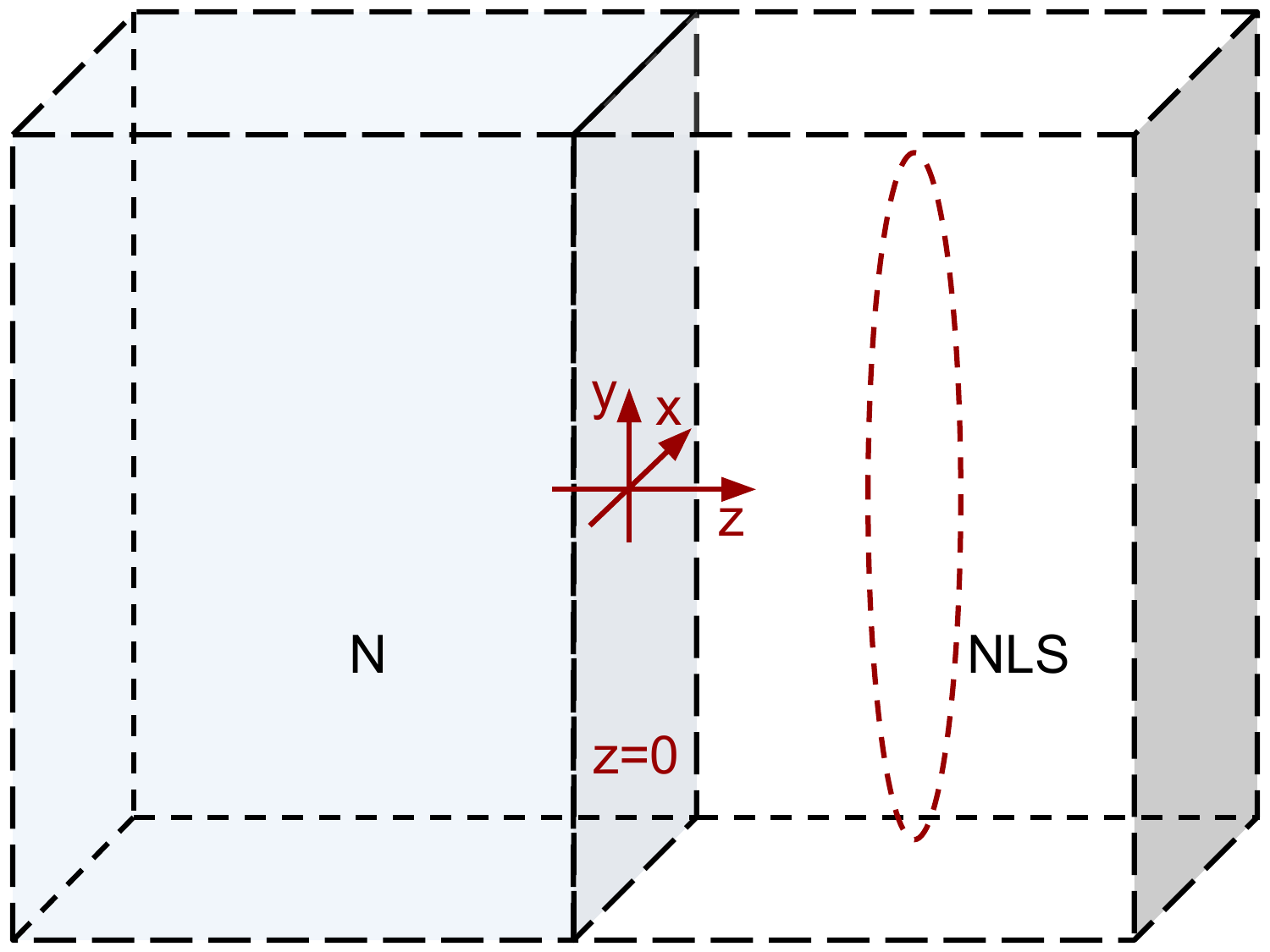}
		\caption{Schematic representation of a N-NLS junction, with the interface parallel to the plane of the nodal line.}
		\label{fig:singleInterface}
	\end{figure}

\begin{figure}
		\includegraphics[height=0.17\textheight]{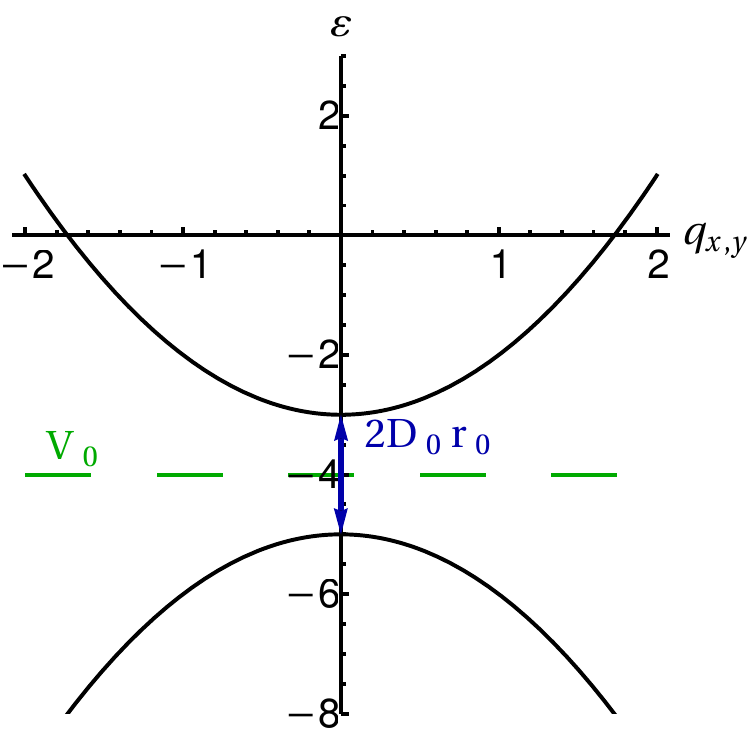}
		\includegraphics[height=0.17\textheight]{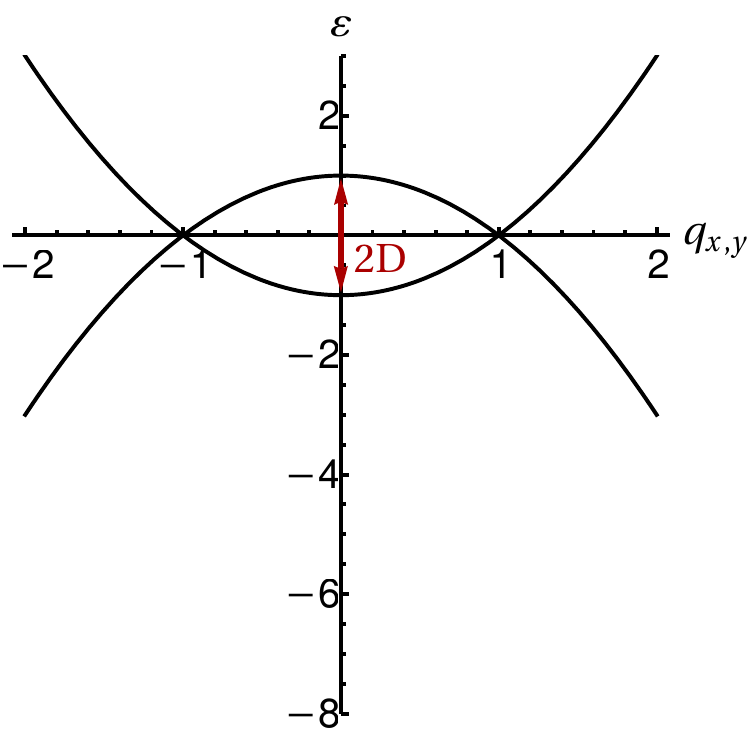}
		\caption{Schematic representation of the band structure in the metallic region $z<0$ (left) and in the semimetallic region $z>0$ (right). \label{fig:bands}
		}
	\end{figure}

	We begin by considering a sharp interface between a normal metal and a nodal-line semimetal, as represented in Fig. \ref{fig:singleInterface}.
	It can be modelled as a special case of parameter discontinuity in the Hamiltonian in Eq. \eqref{eq:H}. We can, in practice, consider this interface by choosing $a(z)=a_0\Theta\left(-z\right)+a\Theta\left(z\right)$, with \mbox{$a_0<0<a$}. A more general choice of function is also possible \cite{Buccheri2024}, but is not expected to affect the main features of the transmission function in this context.
	In addition, we choose a position-dependent band shift in the form  $V(z)=V_0\Theta\left(-z\right)$ and we will focus on the instance $V_0<0$, in which case the carriers in the metallic regions are electrons. Figure \ref{fig:bands} schematically shows the two different quasiparticle spectra on the two sides of the junction. We also define for convenience the quantity $\varepsilon_{0}\equiv \varepsilon-V_0$. 
	Using this quantity, the energy of the bulk excitations on the N side also has the form of the dispersion \eqref{eq:nrg}
	\begin{equation}\label{eq:enormal}
		\varepsilon_{0,\pm}\left(\boldsymbol{q}\right)=\pm\sqrt{q_{z}^{2}+m_{0}^{2}\left(q_\mathrm{p}\right)}\;,
	\end{equation}
	here written in terms of dimensionless variables and using the function
	\begin{equation}\label{eq:m0}
		m_0=m_0\left(q_{p}\right)=D_0\left( q_\mathrm{p}^2 + r_0 \right)\;,
	\end{equation}
	where $r_0=a_0/a$. 
	As translation invariance in the $z$ direction is broken, the corresponding momentum is not conserved anymore. Instead, we will use as quantum numbers the energy of the state and the particle/hole branch, which we denote as $\nu$ and $\nu_0$ in the NLS and in the metallic regions, respectively. The absolute value of the momentum in the $z$ direction as a function of the quantum numbers is
	\begin{align}\label{eq:qq0def}
		q_z&=\sqrt{\varepsilon^2-m^2\left(q_\mathrm{p}\right)}\,,\\ q_0&=\sqrt{\varepsilon_0^2-m_0^2\left(q_\mathrm{p}\right)}\;\nonumber,
	\end{align}
    \buc{which we write here for later use, see Secs. \ref{subsec:transSingel}, \ref{sec:singlerefraction} and \ref{sec:slabtransparent}.}
	
	We study the problem of transmission through the N-NLS interface in two limiting regimes: first the perturbative regime in the tunneling between the surfaces, then the transparent limit. 
	It is important to remark, at this point, that the Fermi surfaces of the two samples have distinct topologies: while in N region it is an ellipsoid, in the NLS region the Fermi surface is a torus, which implies that there are distinct regions in the parameter space. In order to see this, we note that the states that can propagate in the bulk of the NLS are contained in an annulus when projected onto the interface BZ, namely, the two-dimensional projection of the bulk toroidal Fermi surface. This is identified by the condition that the momentum in the $z$ direction \buc{\eqref{eq:qq0def}} is real, reading 
	\begin{equation}\label{eq:qrange}
		1-\frac{\left|\varepsilon\right|}{D}<q_{p}^{2}<1+\frac{\left|\varepsilon\right|}{D}\,,
	\end{equation}
    for $-D<\varepsilon<D$.
	At the same time, the projection of the propagating states in the N region onto the interface BZ is the full disk $q_\mathrm{p}^2<\frac{\left|\varepsilon_0\right|}{D_0}-r_0$. We can therefore identify three regimes: (i) The diameter in the $q_x$$q_y$ plane of the Fermi surface in N is small compared to the inner diameter of the Fermi surface of the NLS. Focusing for definiteness on an electronic Fermi surface, this is possible if
     \begin{equation}\label{eq:V0>smthg}
         -D_0(1+r_0)<V_0<0
         \;,
     \end{equation}
    for energies in the range
    \begin{equation}\label{eq:erange>0}
                0<\varepsilon<D\min\left\{\frac{D_0(1+r_0)+V_0}{D+D_0},1\right\} \;.
    \end{equation}
     In this regime, the absence of matching states implies that the transmission always vanishes exactly. (ii) The transverse size of the Fermi surface of the metal is intermediate between the inner and the outer diameters, in which case the transmission probability is different from zero, but never reaches unit value, as seen in the right panel of Fig. \ref{fig:Tsingle}). (iii) The projection of the N Fermi surface completely covers the projection of the NLS Fermi surface onto the interface BZ. As seen in the left panel of Fig. \ref{fig:Tsingle}, the transmission probability reaches unit value for some values of the momenta and the expected conductance per unit surface is maximal.
     In the specific case $D=D_0$, this is ensured when the inequality
     \begin{equation}\label{eq:V0<smthg}
         V_0<-D_0(1+r_0)\;
    \end{equation}
    is satisfied. We refer the reader to the more general description in App. \ref{app:singletransparent}, where the other parameter regimes are discussed.

	\subsection{Transport through a single interface }\label{subsec:transSingel}
	We start by exploiting the results of Sec. \ref{sec:obc} and consider the two samples as disconnected, but weakly hybridized by the electronic tunneling between the interfaces at $z=0$.
	Here we assume that the transverse momentum $\mathbf{q}_\mathrm{p}$ is conserved by the tunneling process as a consequence of the translational invariance in the plane of the interface. This may be violated in the presence of inhomogeneities, such as charged impurities or vacancies, or if there is a certain degree of roughness at the interface. It is nevertheless a reasonable approximation if the interface is sufficiently clean.
	
	We start from two disconnected materials with the band structure represented in Fig. \ref{fig:bands}: open boundary conditions on each of them are imposed as in Eq. \eqref{eq:bc}, but with two generically different parameters $\alpha$ and $\alpha_0$.
    \buc{The equilibrium chemical potential of the whole system is set at $\mu$. In order to study the current in linear responses, we impose a small imbalance $\delta \mu = e \delta V$ between the two samples.}
 The (elastic) tunneling is described by the term
	\begin{equation}\label{eq:tunneling}
		H_T=\lambda\sum_{\boldsymbol{k}_\mathrm{p}}\psi_{0}^{\dagger}\left(\boldsymbol{k}_\mathrm{p},z=0\right)\psi\left(\boldsymbol{k}_{p},z=0\right) +H.c.
	\end{equation}
	where the subscript "$0$" denotes the metallic lead, infinitely extended in the negative $z$ direction, see Fig. \ref{fig:singleInterface}. In the expression above there appears a tunneling amplitude $\lambda$
	and we assume, for simplicity, that it is approximately constant in a region around the Fermi energy of order $\sim k_BT$. Note that the operators $\psi$ have two components, but the tunneling is diagonal in the internal degree of freedom. 
	
	We first consider the regime $ |\lambda| n_0\ll 1$, where $n_0$ denotes the density of states at the Fermi energy. 
	The current through the interface is the derivative of the number of carriers in the lead $j\left( t\right) =\left\langle \dot N_0\right\rangle $, in the presence of the weak perturbation in Eq. \eqref{eq:tunneling}. The standard problem of perturbative tunneling (see, e.g. \cite{Mahan}), together with the condition of conservation of the transverse momentum, brings the linear-response current per unit surface in the form
	\begin{align}\label{eq:jweak}
		j=&\delta V\frac{e^2\left|\lambda\right|^{2}}{\pi \hbar^3 v^{2}}
		\mbox{Tr}\left\{ \hat{g}_{\alpha}\hat{g}_{\alpha_{0}}\right\}
		\nonumber\\
		&\times
		\intop d\varepsilon
		\intop
		\frac{dq_{p}}{2\pi}q_{p} n_0\left(q_\mathrm{p},\varepsilon \right)n\left(q_\mathrm{p},\varepsilon \right)
		\frac{\beta}{4\cosh^2\frac{\beta\left(\varepsilon-\mu\right)}{2}}
	\end{align}
	where the integration range of $q_\mathrm{p}$ is in Eq. \eqref{eq:qrange} and $\delta V$ is a weak potential imbalance between the two subsystems. The momentum-resolved densities of states in the lead $n_0=dq_0/d\varepsilon$ and in the sample $n=dq_z/d\varepsilon$ are directly obtained from Eq. \eqref{eq:qq0def}, while the matrices $\hat g$ and $\hat g_0$ encode the spinor structure of the surface of the two materials and, together with additional details about the derivation, are provided in Appendix \ref{app:singletransparent}.
	It is worth mentioning that the equation above only sums the states which are transmitted into propagating states, which are detected asymptotically far from the interface. The tunneling into surface states is also possible and produces, after a transient, a surface charge accumulation. While we do not explicitly model this electrostatic barrier here, the net effect is a suppression of the effective tunneling coefficient in Eq. \eqref{eq:tunneling}. Assuming this is already taken into account in the tunneling parameter, one arrives at the low-temperature linear-response conductance per unit surface 
	\begin{equation}\label{eq:sigmawc}
		\sigma=\frac{2e^{2}\left|\lambda\right|^{2}}{\hbar^3\pi\beta v^{2}D}\ln\left(2\cosh\frac{\beta \mu}{2}\right)\cos^{2}\frac{\alpha-\alpha_{0}}{2} \;.
	\end{equation}
    As seen from the expansion of the argument of the logarithm $2\cosh\frac{\beta \mu}{2}\approx 2+O\left(\beta^2\mu^2\right)$, the conductance vanishes linearly in the temperature in the limit ${\beta\mu}\ll1$. Moreover, we observe that the surface parameters appear in a remarkably simple combination, which showcases the dramatic effect of the boundary angle mismatch on the weak-tunneling conductance. 
	
\begin{figure}
		\includegraphics[height=0.15\textheight]{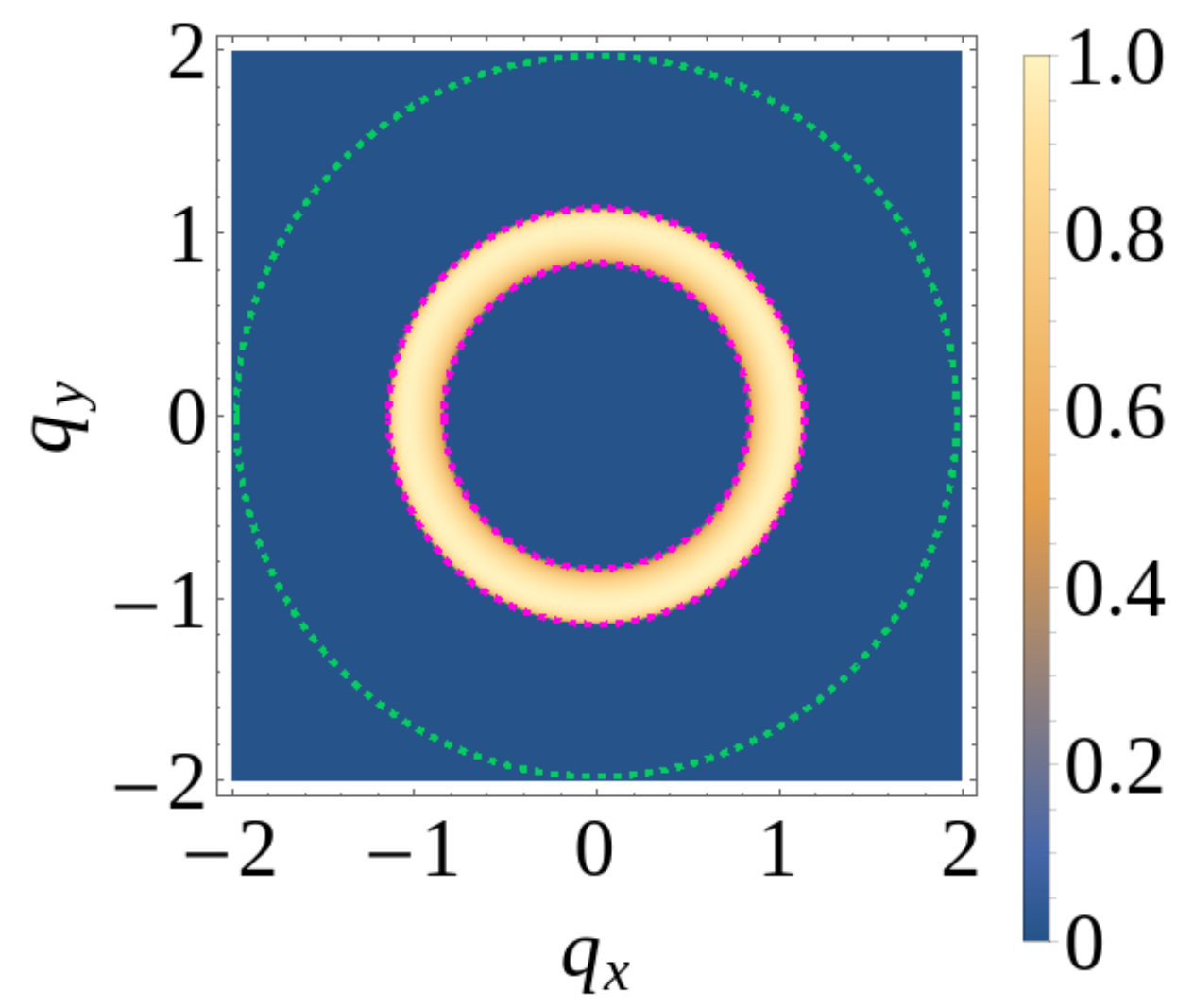}
		\includegraphics[height=0.15\textheight]{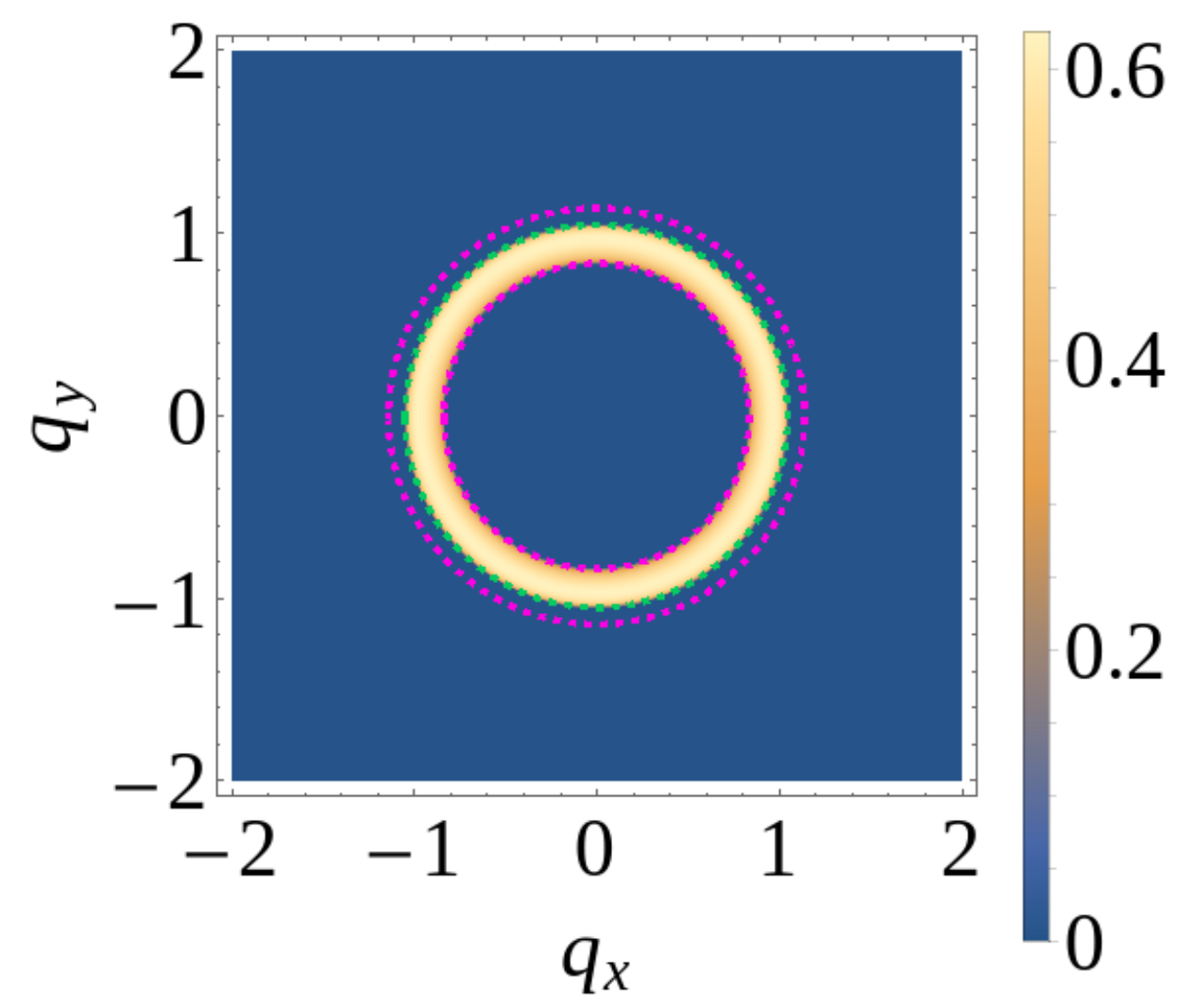}
		\caption{Transmission through a single interface as a function of the momenta in the $xy$ plane. In purple, the edges of the Fermi surface of the NLS, in green, those of the metallic region. The parameters chosen here are $\varepsilon=0.3$, $D=D_0=1$, $r_0=0.1$, $V_0=-3.5$ (left) and $V_0=-0.7$ (right). \label{fig:Tsingle}}
	\end{figure}

	Conversely, the details of the boundary do not matter in the strong tunneling regime $|\lambda| n_0 \gg 1$. Within our assumption of ballistic transport, this regime can be tackled as a quantum-mechanical scattering problem by matching the single-particle wavefunctions at the interface. Details are presented in App. \ref{app:singletransparent}, while here we just state the result for the transmission probability as a function of the energy and the transverse momentum of the incoming electron or hole
	\begin{equation}\label{eq:Tsingletransparent}
		\mathcal{T}_1\left( \varepsilon,q_\mathrm{p}\right)= \frac{4q_{0}q_{z}\left|\varepsilon-m\right|\left|\varepsilon_0-m_{0}\right|}{\left(q_{0}\left|\varepsilon-m\right|+q_{z}\left|\varepsilon_0-m_{0}\right|\right)^{2}}.
	\end{equation}
	We note that this function describes both the scenarios in which an incoming electron is transmitted as an electron (normal tunneling) {if $\varepsilon>0$} and that in which it is transmitted as a hole (Klein tunneling) {if $\varepsilon<0$}. 
	In Figure \ref{fig:Tsingle}, we illustrate the transmission probability \eqref{eq:Tsingletransparent} at given energy as function of the transverse momenta. In the left panel, the chosen parameters fall in the regime (iii) of Sec. \ref{subsec:intmodel} and we observe that that the perfect transmission ($\mathcal{T}_1 = 1$) is reached when
	\begin{equation}\label{eq:q0}
	q_\mathrm{p}=\sqrt{\frac{D\varepsilon_0-r_0D_0\varepsilon}{D\varepsilon_0-D_0\varepsilon }} \,.
	\end{equation}
	This is in contrast to the right panel of Fig. \eqref{fig:Tsingle}, in which the parameters fall in the regime (ii) and the transmission function in Eq. \eqref{eq:Tsingletransparent} never reaches unit value. 
 
	We will continue the analysis of transport in the transparent interface limit in section \ref{sec:N-NLS-N} and consider now the refraction of the electron trajectories originated by the change in the dispersion through the single interface.
	
	\subsection{Refraction and electronic optics}\label{sec:singlerefraction}
	
	As already discussed in Sec. \ref{sec:intro}, topological semimetals are good candidates for observing phenomena connected to the geometric electron optics, such as focusing.
	We show now that this is achievable because of the bulk band dispersion alone, and in particular because of the presence of a nodal line around the $\Gamma$ point, without the need of an applied electrostatic potential or electric field.
	
	Consider an electron incident on the interface at $z=0$ from the N side, with energy $\varepsilon$ and angle $\theta_n$ with respect to the normal. The electron is transmitted to the NLS side with probability given by Eq. \eqref{eq:Tsingletransparent} and with exit angle $\theta_{s}$. 
	The components of the semiclassical velocity $\boldsymbol{u}$ in the NLS sample can be parametrized by its modulus $u$ and the angle $\theta_s$ with the normal to the surface
	\begin{equation}\label{eq:us}
		u_{p}\equiv u\sin\theta_{s}\;,\qquad
		u_{z}\equiv u\cos\theta_{s}\;,
	\end{equation}
	while the angle in the plane of the interface is irrelevant, due to the cylindrical symmetry of our problem. Analogously, the velocity $\boldsymbol{w}$ in the metallic sample can be parametrized by its module $w$ and the incoming angle $\theta_n$
	\begin{equation}\label{eq:ws}
		w_{p}\equiv w\sin\theta_{n}	\;,\qquad
		w_{z}\equiv w\cos\theta_{n}\;.
	\end{equation}
	Following the procedure detailed in App. \ref{app:angles}, we arrive at the generalized Snell's law
	\begin{equation}\label{eq:Snell}
		\tan\theta_{s}	= \chi\left(\varepsilon\right)  \left(\tan\theta_{n}-\tan\theta_{n}^{*}\right)\;,
	\end{equation}
	with $\chi$ a function of the energy and of the material parameters
	\begin{equation}\label{eq:f}
		\chi\left(\varepsilon\right) = \frac{D^{2}}{D_0^{2}\varepsilon}
		\frac{2  q_{0}^3(1)}{ 2\varepsilon_0^{2}+\left(1+r_{0}\right)q_0^2(1)}\,,
	\end{equation}
	having used $q_0(1)=\sqrt{\varepsilon_{0}^{2}-D_0^{2}\left(1+r_{0}\right)^{2}}$, the latter being the momentum defined in Eq. \eqref{eq:qq0def} with $q_\mathrm{p}=1$. Note that the above expression has a different sign when the energy is above or below the nodal line.
	The angle $\theta_{n}^*$, such that
	\begin{equation}\label{eq:thetastar}
		\tan\theta_{n}^{*}=2\frac{D_0^{2}\left(1+r_{0}\right)}{q_0(1)}\,,
	\end{equation}
	identifies an incoming electron which is transmitted exactly on the nodal line, i.e., with an incidence angle such that the electron exits normally to the interface on the NLS side, which takes place when the transverse momentum matches the nodal line.
	Interestingly, Eq. \eqref{eq:Snell} shows that the exit angle $\theta_{s}$ changes sign when the incidence angle moves across the value $\theta_{n}=\theta_{n}^{*}$.
	
\begin{figure}
		\includegraphics[height=0.18\textheight]{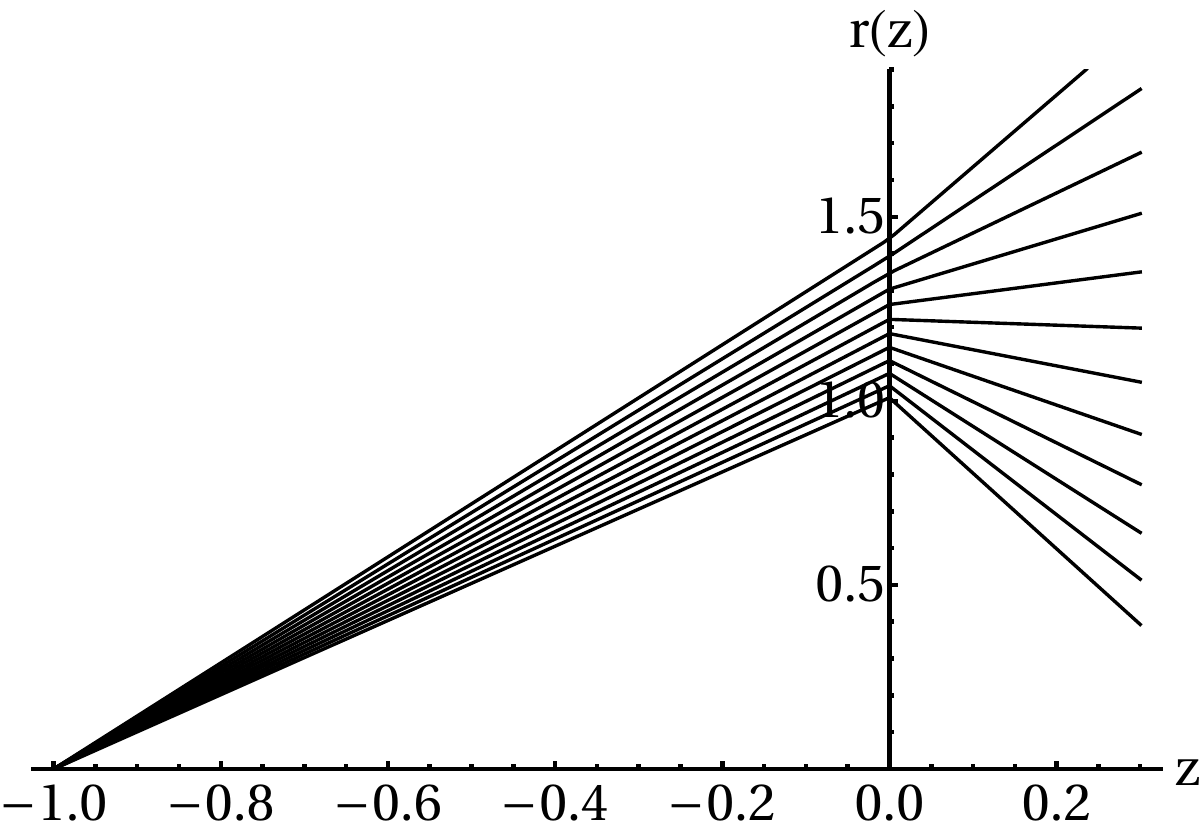}		\includegraphics[height=0.18\textheight]{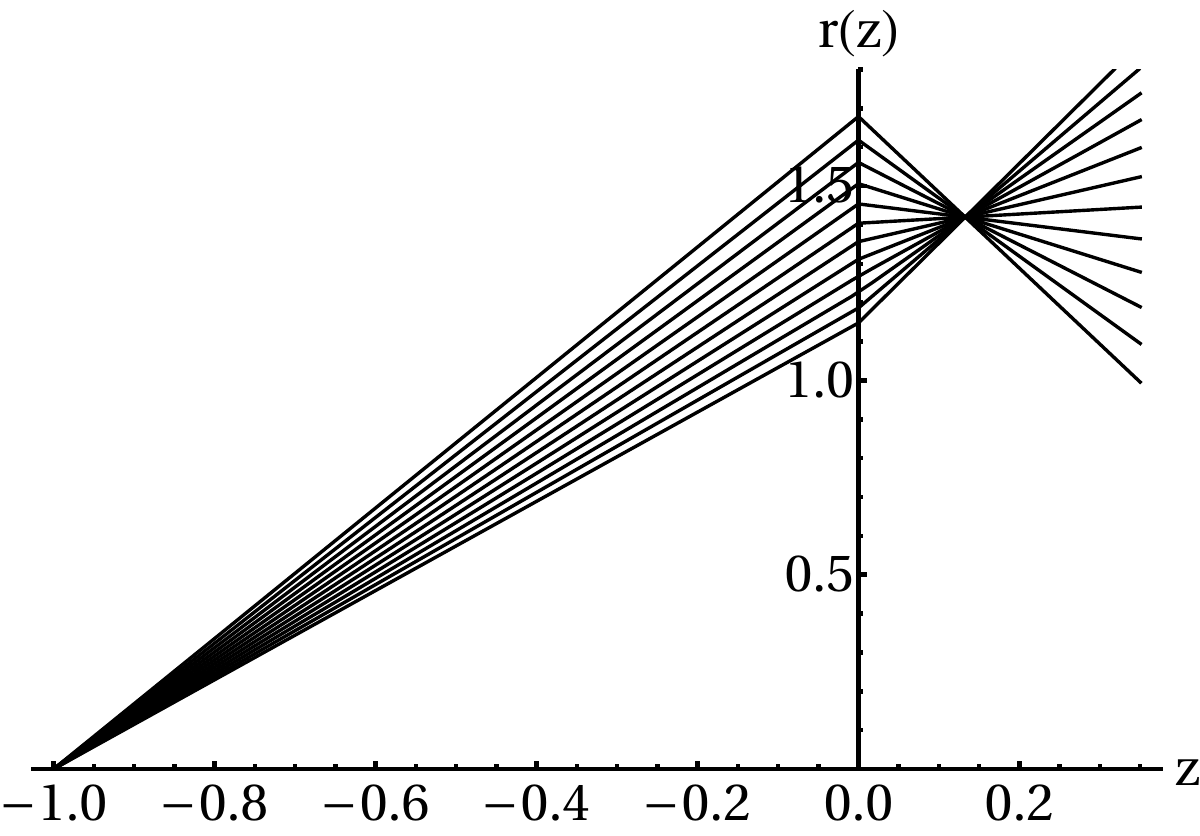}
		\caption{Refraction of monochromatic electrons from a N-NLS interface on a ring in real space. The trajectories of the transmitted electrons from a point source are plotted for various angles at $V_0=-2$, $D=D_0=1$, $r_0=1$. Top: normal transmission $\varepsilon=0.1$. Bottom: Klein transmission $\varepsilon=-0.1$. Only the trajectories with nonzero transmission amplitude are shown. \label{fig:focussingle}}
	\end{figure}

	Imagining a localized source of electrons as in Fig. \ref{fig:focussingle} and defining an axis as the line passing through the source and perpendicular to the interface, several scenarios can arise, according to the doping level. For $\varepsilon>0$, the change in the sign of the velocity implies that the electrons with transverse momenta enclosed by the nodal ring invert the component of their velocity parallel to the interface. In other words, electrons with $q_\mathrm{p}<1$ start traveling back toward the axis after crossing the interface, without ever changing the components of their momentum parallel to the interface: for normal (electron-electron) transmission, the refraction index is effectively negative for these states, see Fig. \ref{fig:focussingle}. The phenomenon described here is different from what is observed in doped graphene \cite{Cheianov2007} in two aspects: first, the change in the sign of the velocity is not originating from the electron being transmitted as a hole; second, refraction indices of opposite sign coexist in different parts of the BZ.
	
	For $\varepsilon<0$, the electron is transmitted as a hole (Klein tunneling). The exit angle is still given by Eq. \eqref{eq:Snell}, but the function \eqref{eq:f} is negative \cite{Buccheri2022b}. In this situation, as shown in Fig. \ref{fig:focussingle}, the transmitted rays focus on a ring in real space.
	Let us consider a point source of electrons at distance $d_s$ from the interface on the metallic side: electrons arrive with a given energy and various incoming angles on the interface, such that the transverse momentum $q_\mathrm{p}$ is within the range in Eq. \eqref{eq:qrange}. The latter requirement implies that the transmission probability is non-vanishing in the regime (iii) and can be directly translated into a range of incidence angles for which transmission into the NLS is possible, see Eq. \eqref{eq:thetarange} and the discussion in Appendix \ref{app:angles}.
	The distance from the axis of the electron/hole trajectory can be written in terms of the incidence angle and the $z$ coordinate 
	\begin{equation}
		r\left(z\right) = \begin{cases}
			\left( d_s+z\right)  \tan \theta_n & z<0 \\
			d_s\tan\theta_{n} + z \tan \theta_s & z>0
		\end{cases}.
	\end{equation}
	From the path above, one sees that for $\tan \theta_s<0$ there is a ring of radius $R=d_s\tan\theta_n^*$ at distance \mbox{$d_R$} from the interface, in the semimetallic region, with \mbox{$d_R=d_s/|\chi\left(\varepsilon\right)|$}. As this is a direct consequence of the presence of a nodal line, it can be interpreted as a way of observing it in real space. We will discuss an experimental setup to probe this effect in Sec. \ref{sec:N-NLS-N}.

    \buc{As mentioned in Sec. \ref{sec:NLS}, a more general shape of the nodal line can be taken into account by an angle-dependent rescaling of the momenta. The phenomenon of negative refraction for momenta inside the nodal line is originated only from the sign change of the velocity across the interface, which is guaranteed by the existence of a nodal line, and therefore, it will be present regardless of its shape. In other words, both panels of Fig. \eqref{fig:focussingle} will be qualitatively the same. However, we expect the focusing effect for $\varepsilon<0$ to take place at a distance $d_R$ which becomes angle-dependent. If the nodal line is reasonably close to a circle, we still expect to detect a focusing pattern that displays the surviving discrete rotational symmetries, possibly blurred due to the deformation.}

	\section{N-NLS-N heterostructure}\label{sec:N-NLS-N}
	
	The problem of a double interface is similar to the one considered above \cite{Fu2022}. We take a slab of NLS and insert two interfaces with a metallic sample, which we assume fully transparent for simplicity. As in Sec. \ref{sec:N-NLS}, the interfaces are parallel to the plane of the nodal line, at $z=-\frac{L}{2}$ and  $z=\frac{L}{2}$.
	In order to gain insight on the kind of states that can carry charge and energy through the system, we first consider an isolated slab and show that there is a class of states resulting from the hybridization of the drumhead states on the two opposite surfaces. We dub them "drum"  states. We then study the transport through the slab in Secs. \ref{sec:slabtransparent} and \ref{sec:electrict} and highlight the effect of electron refraction from the nodal line in Sec. \ref{sec:slabrefraction}.
	
	\subsection{Spectrum in a slab}\label{sec:slab}
	We start our analysis from the spectrum of a NLS slab, in particular by studying the effect of the boundary parameter $\alpha$, by deriving the drum states and the width in which they exist.
	We impose the boundary condition in Eq. \eqref{eq:bc} at $z=-L/2$ and we consider the class of boundary conditions that do not spoil the defining mirror symmetry in Eq. \eqref{eq:mirror}. As the symmetry is implemented by the operator $\tau_z$, the boundary condition to be imposed at the surface $z=L/2$ is 
	\begin{equation}\label{eq:otherBC}
		B\left(-\alpha\right) \psi \left(\frac{L}{2}\right) = \psi\left(\frac{L}{2}\right).
	\end{equation}
	Following, e.g., \cite{Buccheri2022b}, one arrives at the quantization equation
	\begin{equation}\label{eq:slabquantization}
		\tan\left(q_zL\right)=\frac{q_z\sin\alpha}{\varepsilon\cos\alpha-m}\,,
	\end{equation}
	in which {$q_z$ is given in Eq. \eqref{eq:qq0def}}
    and $m$ is the function \eqref{eq:m} of the transverse momentum, as defined in Eq. \eqref{eq:m}. 
	The boundary conditions in \cite{Burkov2011} correspond to the choice $\alpha=\pi/2$, for which the "pseudo-spins" on opposite surfaces are orthogonal. Equation \eqref{eq:slabquantization} is able to capture the finite-size effects on the spectrum of a NLS slab for general mirror-symmetry-preserving boundary conditions.
	
	For generic values of the boundary angle $\alpha$, we are not aware of analytic solutions to Eq. \eqref{eq:slabquantization}, which is therefore tackled numerically. Analytic solutions are only obtained in specific limits, as discussed in App. \ref{app:quantization}.
	Interestingly, Eq. \eqref{eq:slabquantization} can be directly continued to imaginary values of the momentum, see Eq. \eqref{eq:slabquantloc}: as a matter of fact, there exist solutions with \mbox{$q_z=i\kappa$}, where $\kappa=\sqrt{m^2-\varepsilon^2}$ takes the meaning of the inverse of a penetration depth. Such solutions represent drum states, arising from the hybridization of the drumhead states on opposite surfaces and have therefore nonvanishing weight on \emph{both} surfaces. They are labeled by the transverse momentum $q_\mathrm{p}$ and the parity under the mirror reflection in Eq. \eqref{eq:mirror}.

\begin{figure}
		\includegraphics[height=0.18\textheight]{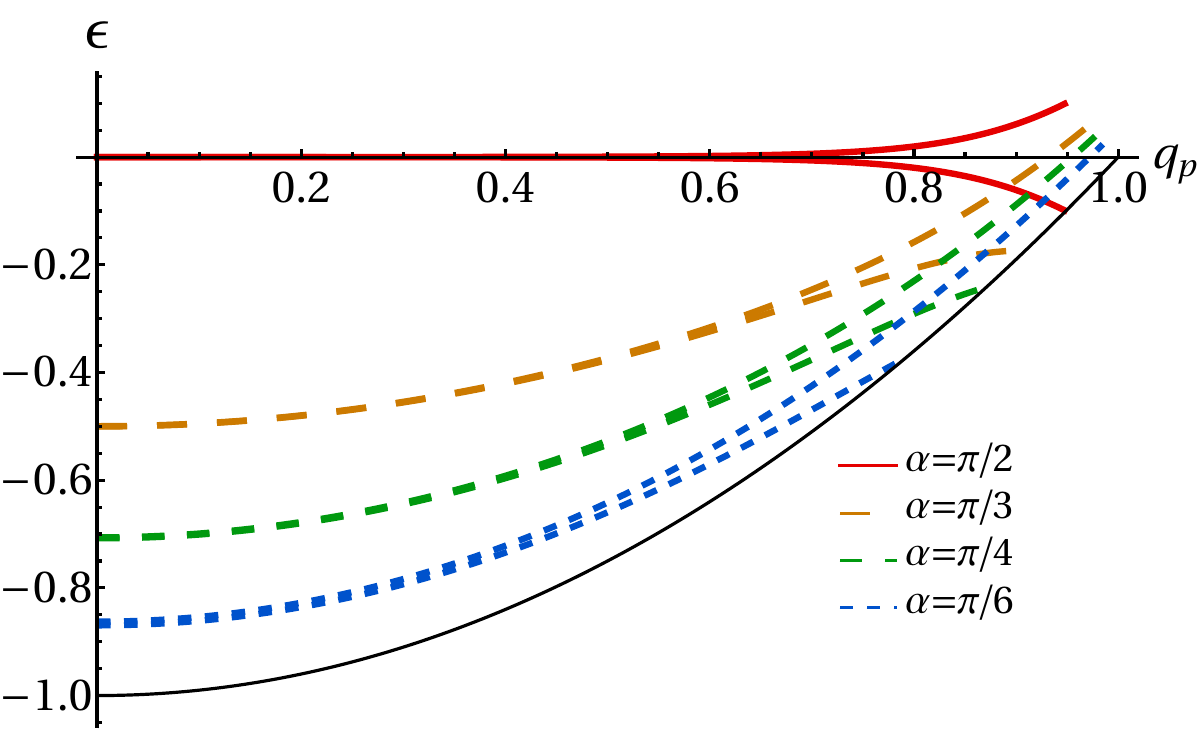}
		\caption{Dispersion of the surface states as a function of the transverse momentum for $L=10$ and various values of the boundary angle, from the numerical solution of Eq. \eqref{eq:slabquantization}. At finite size, each doublet splits into a pair of drum states, as evident on the right part of the graph.
			 The solid black line $\varepsilon=m$ is the threshold of continuum states.
        The splitting is larger as one approaches the boundary of the support because of the diverging penetration length, which determines a larger overlap between the drumhead states.
    \label{fig:L10energy}}
        
	\end{figure}

\begin{figure}
		\includegraphics[height=0.18\textheight]{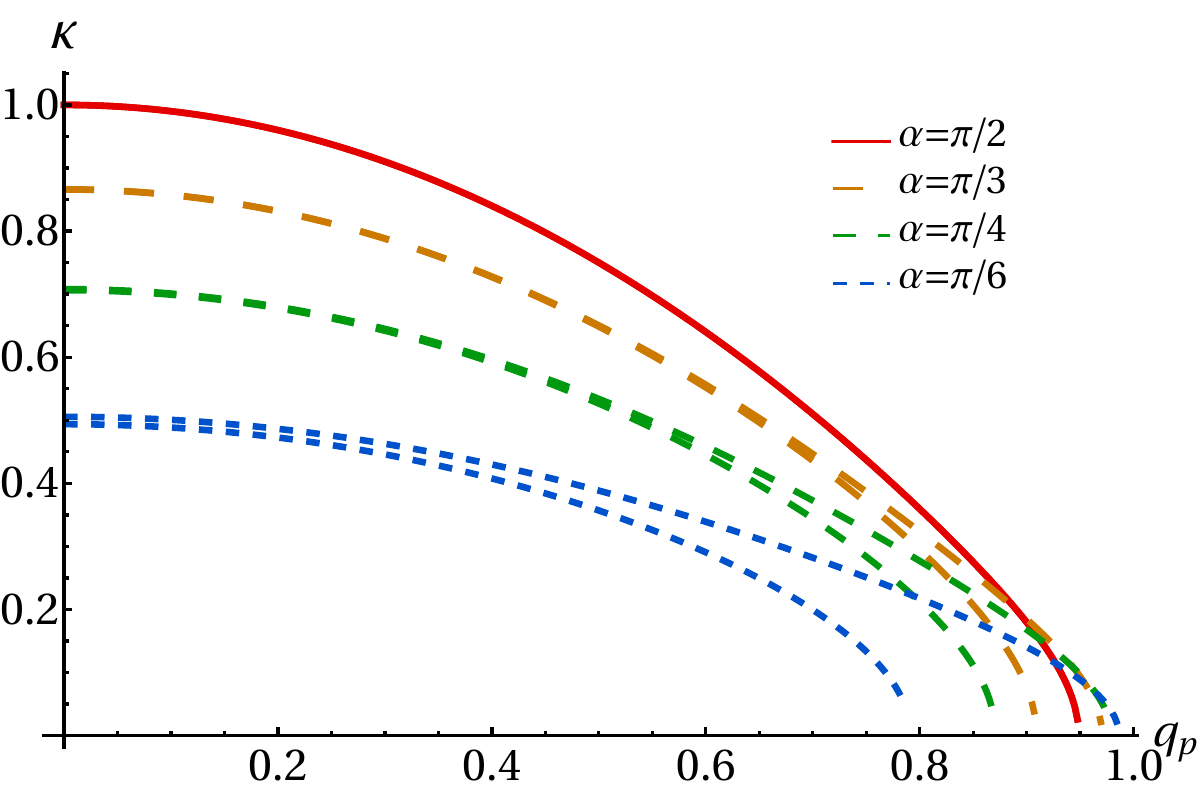}
		\caption{Inverse penetration depth of surface states as a function of the transverse momentum for $L=10$ and various values of the boundary angle.
			 \label{fig:L10kappa}}
	\end{figure}

	In the large-slab limit, and in particular in the regime $L|m|\sin\alpha\gg1$, one recovers the states \eqref{eq:drumheadstate}, with the inverse penetration length given in Eq. \eqref{eq:kappainf} and the dispersion in Eq. \eqref{eq:ssdispersion}. In Figures \ref{fig:L10energy} and \ref{fig:L10kappa}, we show the numerical solution of Eq. \eqref{eq:slabquantization} for a moderately large size. A completely flat band exists only for \mbox{$\alpha=\pi/2$} in the limit \mbox{$L\to\infty$}, see Eq. \eqref{eq:ssdispersion}. For finite size, instead, the hybridization between states on opposite surfaces generates a pair of dispersive bands with opposite eigenvalues under inversion.
	Numerical analysis, see Fig. \ref{fig:L10energy}, shows that one band corresponding to surface states curves upward and is positive around the edge of the support $\varepsilon=-m$. The bottom band remains instead negative and merges into bulk states when $\varepsilon=m$.
	

	\subsection{Transport through a NLS slab}\label{sec:slabtransparent}
\begin{figure}
		\centering
		\includegraphics[height=0.18\textheight]{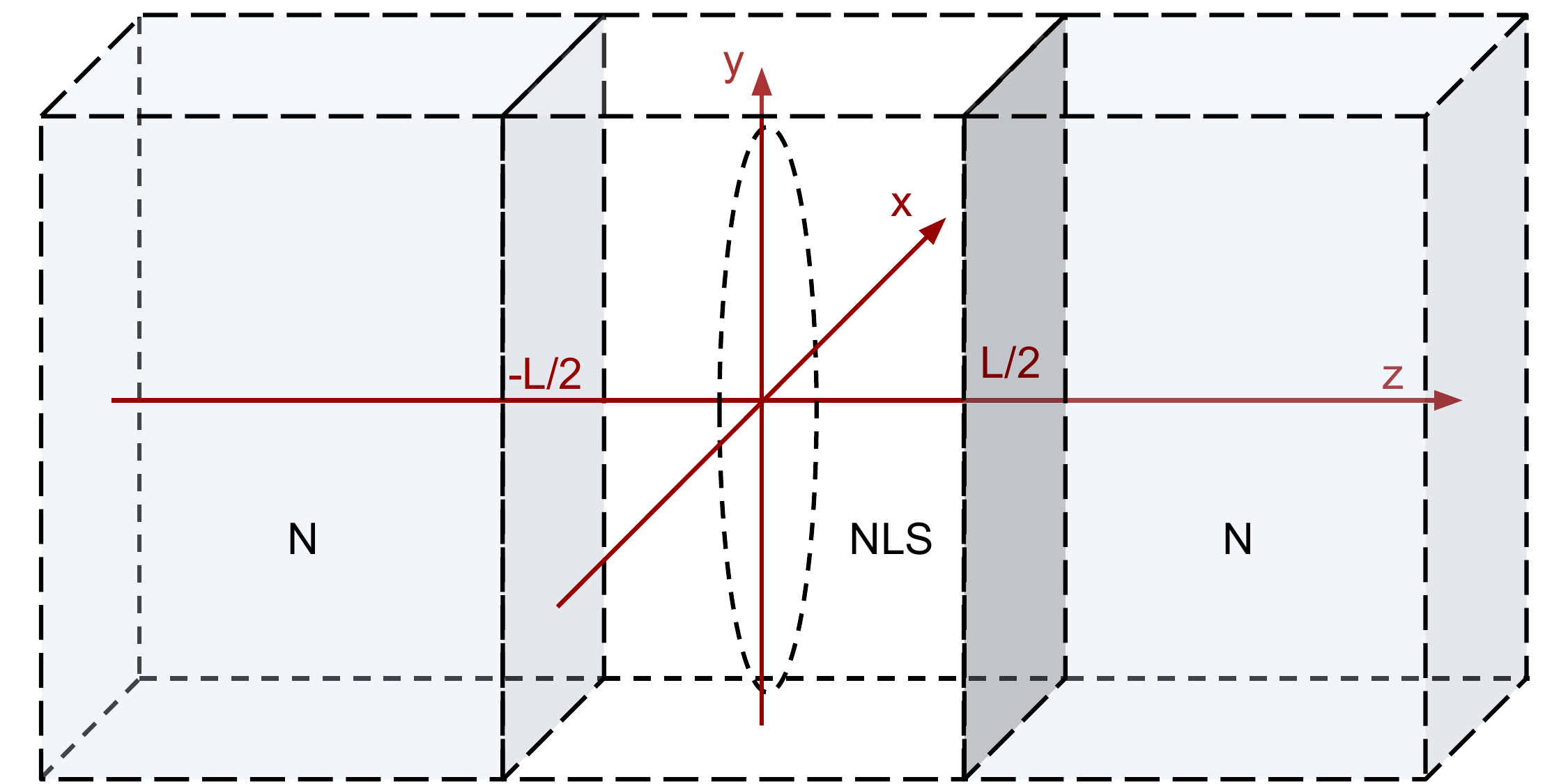}
		\caption{Scheme of a N-NLS-N junction, infinitely extended in the $x$ and $y$ directions. The metallic leads are at $z<-L/2$ and $z>L/2$ and the nodal line is in the plane parallel to the interfaces. } 
		\label{fig:NNLSN}
	\end{figure}

	In this section, we return to the transport analysis from where we left it in Sec. \ref{subsec:transSingel} and connect the NLS slab to two infinite metallic leads, in the configuration of Fig. \ref{fig:NNLSN}. 
    Let us focus, for definiteness, on regime (iii) and energies $-D<\varepsilon<D$.
	By considering the problem of scattering through the double interface, we derive the transmission probability
	\begin{equation}\label{eq:Tslab}
		\mathcal{T}\left(\varepsilon,q_\mathrm{p}\right) =  \frac{q_{0}^{2}q_{z}^{2}}{q_{0}^{2}q_{z}^{2}\cos^{2}\frac{\varphi_\varepsilon}{2}+\left(\varepsilon_{0}\varepsilon-m_{0}m\right)^{2}\sin^{2}\frac{\varphi_\varepsilon}{2}} \,,
	\end{equation}
	where $q_z$, $q_0$ are defined in Eq. \eqref{eq:qq0def} and
	\begin{equation}\label{eq:phase}
		\varphi_\varepsilon\equiv 2q_zL
	\end{equation}
	is the phase acquired by an excitation with energy $\varepsilon$ while completing a back-and-forth path between the two interfaces. The details of the calculation are presented in App. \ref{app:slabtransparent}. We note that there is a series of resonances when this phase is an integer multiple $j$ of $2\pi$, for which perfect transmission is achieved.
	The corresponding energies are 
	\begin{equation}
		\varepsilon_j\left(q_\mathrm{p}\right) = \sqrt{\frac{\pi^2j^2}{L^2}+m^2\left(q_\mathrm{p}\right)}
	\end{equation}
	for integer values of $j$. Following the geometric considerations in Sec. \ref{sec:singlerefraction}, we conclude that, at given energy, there are a series of \emph{resonant angles} $\theta_{n}^{j}$, which correspond to perfect transmission via each of these quantized resonant states. Exploiting the relation between the incidence angle and the transverse momentum, see Eq. \eqref{eq:deltakp}, we conclude that 
	\begin{equation}
		\tan\theta_{n}^{j}\left(\varepsilon\right) = \tan\theta^{*}_n+\frac{2D\sqrt{\varepsilon^2-\frac{\pi^2j^2}{L^2}}}{\varepsilon\chi\left(\varepsilon\right)}
		\,,
	\end{equation}
	with $\theta_n^*$ defined in Eq. \eqref{eq:thetastar}. This expression holds for energies above the first resonance threshold $\pi/L$ and angles within the range specified in \eqref{eq:thetarange}.
	
\begin{figure}
		\includegraphics[height=0.18\textheight]{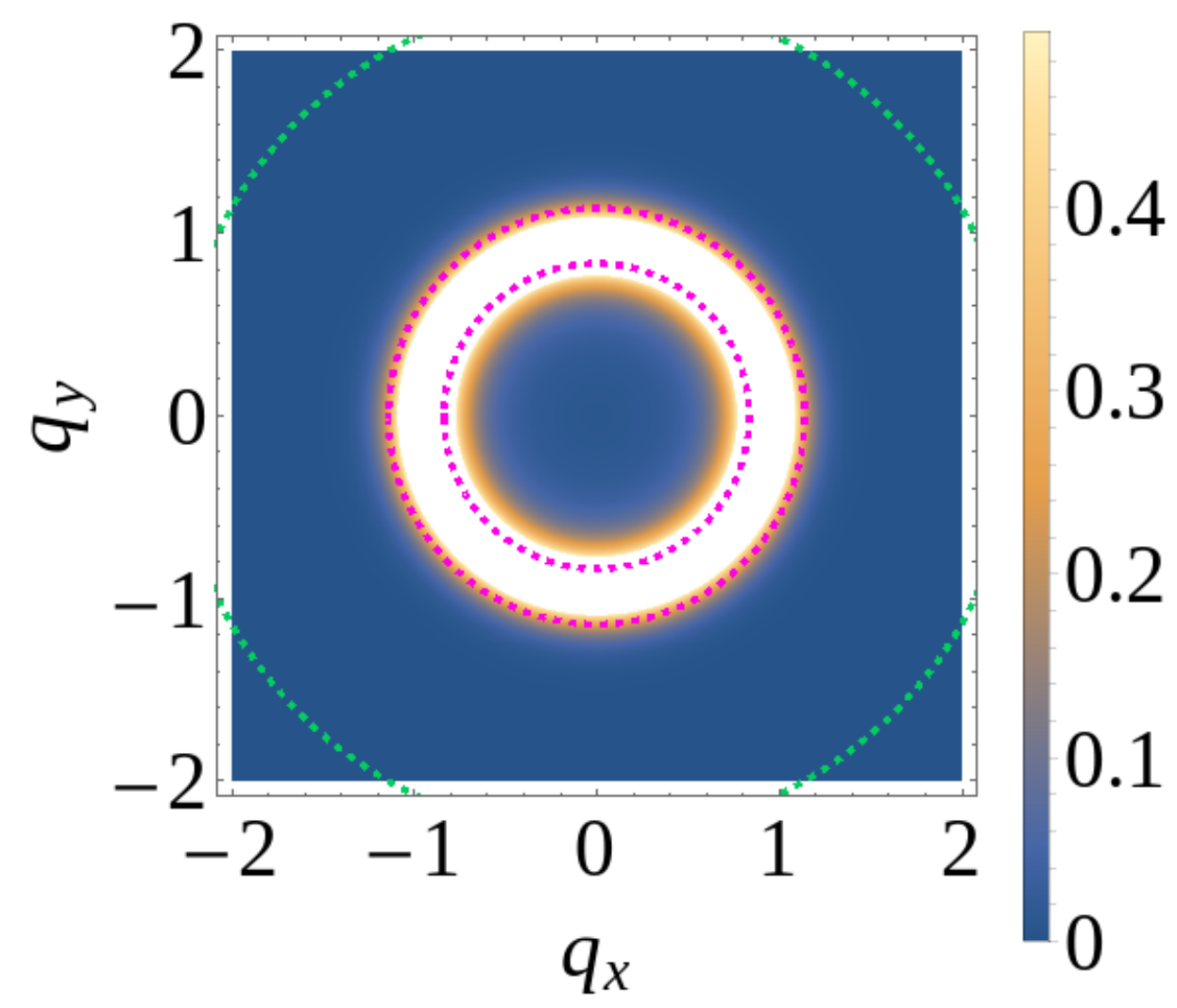}
		\caption{Transmission probability \eqref{eq:Tslab} (real $q_z$) and \eqref{eq:TslabI} (imaginary $q_z$) in the transparent interface limit. The green line denotes the boundary of the projection of the bulk Fermi surface of N onto the interface BZ, while the dashed purple lines denote the boundaries of the projection of the bulk toroidal Fermi surface of the NLS. The states inside the inner circle cannot propagate in the bulk: instead, the non-zero value of $\mathcal{T}$ is originated from the drum states, see Eq. \eqref{eq:TslabI}. The parameter chosen in this figure are $L=3$, $\varepsilon=-0.3$, $V_0=-5.5$, $V=0$, $D=D_0=1$, $r_0=0.1$.\label{fig:Tslab}}
	\end{figure}

	Another important feature of this function is that it can be analytically continued to imaginary values of the momenta $q_z= i\kappa$, obtaining the result
	\begin{equation}\label{eq:TslabI}
		\mathcal{T}\left(\varepsilon,q_\mathrm{p}\right) = 
		\frac{q_{0}^{2}\kappa^{2}}{q_{0}^{2}\kappa^{2} \cosh^{2}\left(\kappa L\right) +\left(\varepsilon_{0}\varepsilon-m_{0}m\right)^{2} \sinh^{2}\left(\kappa L \right)} \;.
	\end{equation}
	As the product $\kappa L$ appears in the hyperbolic functions in the denominator, the expression for transmission through bound states is significantly different from zero only for states close to the nodal line, where the penetration length is comparable to $L$. However, as shown in Fig. \ref{fig:Tslab}, localized states can contribute in the momentum region where no propagating states are allowed, thus, enhancing the conductance at low temperatures.
    A sizeable contribution from the localized states is obtained when the hyperbolic functions in the denominator of Eq. \eqref{eq:TslabI} are at most $O(1)$. Since $\kappa$ is at most $O(D)$, as seen in Fig. \ref{fig:L10kappa} and from Eq. \eqref{eq:kappainf}, this occurs provided $DL\ll1$.
	We note that in this model \cite{Burkov2011,Chan2016a}, the regions inside and outside the nodal line can both host surface states. While for the open boundary condition in Sec. \ref{sec:obc} the boundary parameter selects the support of the surface states, for a transparent interface, the energy $\varepsilon$ will determine in which momentum region evanescent states are present in the junction. The addition of higher-order terms in the low-energy Hamiltonian cures the ambiguity \cite{Shen2017}, but, while turning the analytical calculations more cumbersome, does not add qualitatively new features.

	\subsection{Electric transport}\label{sec:electrict}
	
	We now address the conductance of the slab, focusing in particular on the role of the drum states. We assume that the equilibrium distribution of the carriers in the leads is
	\begin{equation}\label{eq:f0}
		f_0=\frac{1}{e^{\beta\left(\varepsilon_0-\mu\right)}+1}\;,
	\end{equation}
	and is controlled by contacting the sample with voltage-biased reservoirs. Consistently with the choice to measure temperatures in the energy units of Sec. \ref{sec:N-NLS}, we have introduced the notation $\beta\equiv\hbar v k_\mathrm{NL}/k_BT$ for a dimensionless inverse temperature \buc{and $\mu$, the chemical potential, is also in units of $\hbar v k_{\mathrm{NL}}$}. We bias the two leads with a small chemical potential difference, and assuming a low impurity concentration and small enough electron-phonon coupling, we study the regime of coherent transport in linear response.
	
\begin{figure}
		\includegraphics[height=0.18\textheight]{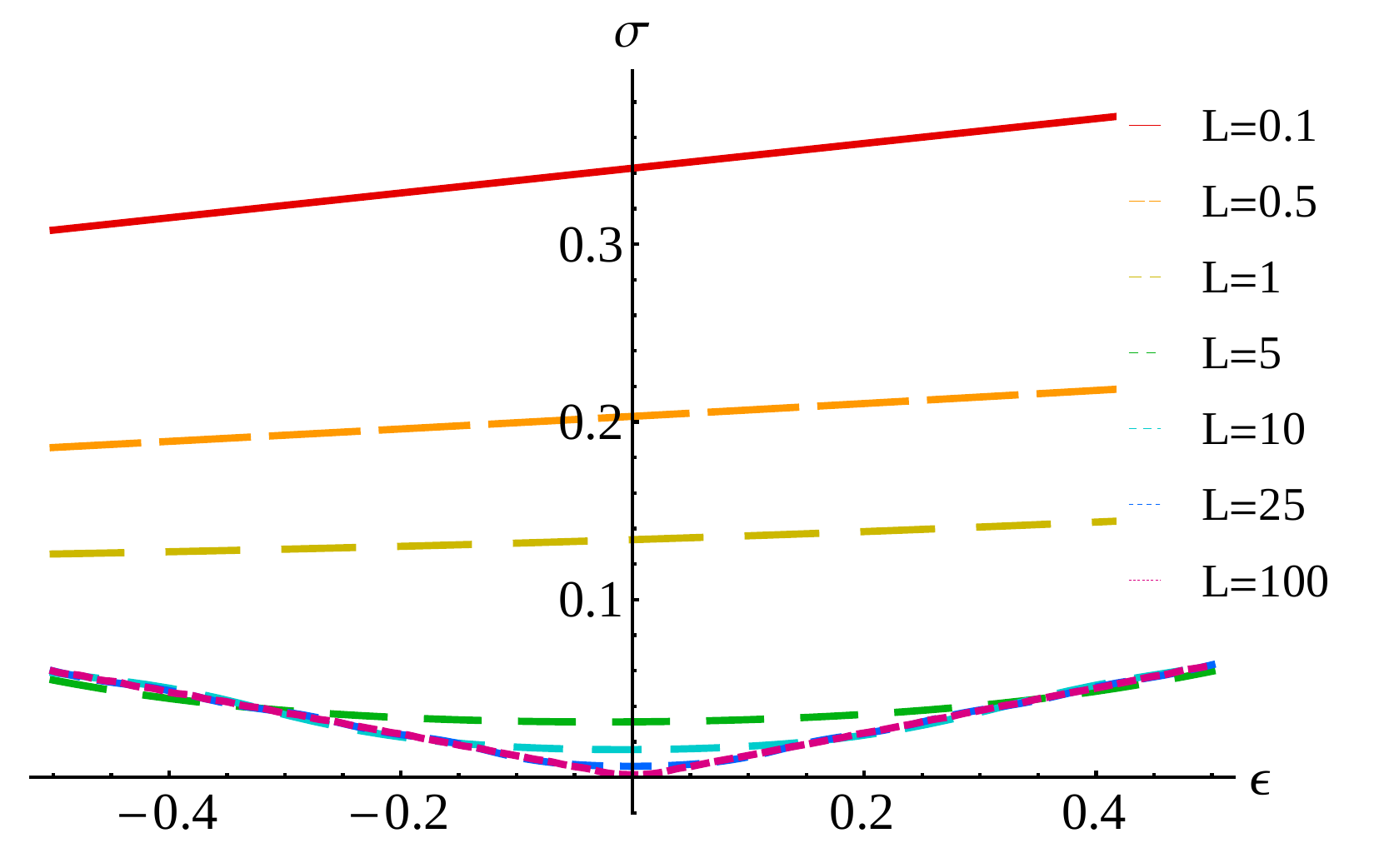}
		\caption{
			Slab thickness dependence of the charge conductance per unit surface in the low-temperature limit according to Eq. \eqref{eq:sigmalow} (in units of $e^2vk_\mathrm{NL}^3/2\pi$) as a function of the energy (in units of $\hbar v k_\mathrm{NL}$).
			The parameters are $D=D_0 =1$, $r_0=0.1$, $V_0=-5$. \label{fig:sigmalow}
		}
	\end{figure}

	With the leads in a metallic regime, the magnitude of $|V_0+\mu|$ is fixed by the Fermi energy of the metal, which is generally larger than the cutoff $\sim D$ discussed in Sec. \ref{sec:NLS} \footnote{For definiteness, $E_0\approx 2.3*10^3K$ in $\mbox{Ca}_3\mbox{P}_2$.}. Since our low energy Hamiltonian is a good description of the system for $\beta D \gg 1$, then in turn $\beta |V_0+\mu| \gg1$. We conclude that, within the regime of validity of our effective description, the system is in the low-temperature regime and the distribution \eqref{eq:f0} is step-like up to exponential corrections. 
    Therefore, we focus on the low-temperature regime and apply the Landauer formula, which yields
	\begin{equation}\label{eq:sigmalow}
		\sigma = \frac{e^2vk_\mathrm{NL}^3}{2\pi} \mathcal{T}\left(\varepsilon\right) +O\left(T^2\right) \;,
	\end{equation}
	in which
	\begin{equation} \label{eq:T(e)}
		\mathcal{T}\left(\varepsilon\right)=\intop \frac{d^2\boldsymbol{q}_\mathrm{p}}{\left(2\pi\right)^2}\mathcal{T}\left(\varepsilon,q_\mathrm{p}\right)
	\end{equation}
	is the total transmission probability at given energy \buc{$\varepsilon$}. 
	Our first observation is that this quantity contains the nodal line radius $k_\mathrm{NL}$, i.e., it can be useful to predict a trend when comparing the conductance of different materials. The transmission function itself is plotted in Fig. \ref{fig:sigmalow}.
    \buc{At $\varepsilon=0$, we note that the bulk contribution to conductance vanishes because of the vanishing density of states.}
	In the limit $L\gg1$, the conductance vanishes linearly for $\varepsilon\to0$. However, at finite size, the zero-temperature conductance shows signatures of the localized drum states: indeed, they turn this quantity finite even for a vanishing bulk density of states.
    Numerical evaluation shows that the zero-temperature conductance dies out as $\sigma\propto 1/\buc{L}$ for large width, despite the exponential localization discussed in Sec. \ref{sec:NLS} and \ref{sec:slab} above.
    \buc{The mechanism behind this result is the divergence of the penetration length of the surface states when approaching the nodal line. The argument is that, while at fixed momentum $q_p$, the overlap between localized states is exponentially suppressed $\sim e^{mL}$, the function $m\left(q_p\right)$ given at the end of Sec. \ref{sec:NLS} tends to zero at the nodal line $\sim q_p^2-1$. When computing the contribution from the whole drumhead state, the resulting integral over $q_p$ goes to zero indeed as $1/L$. }
	We note in passing that the low-temperature electronic thermal conductance is related to the charge conductance \eqref{eq:sigmalow} by the Wiedemann-Franz law and shows similar features.

	\subsection{Current profile}\label{sec:slabrefraction}
\begin{figure}
		\includegraphics[height=0.18\textheight]{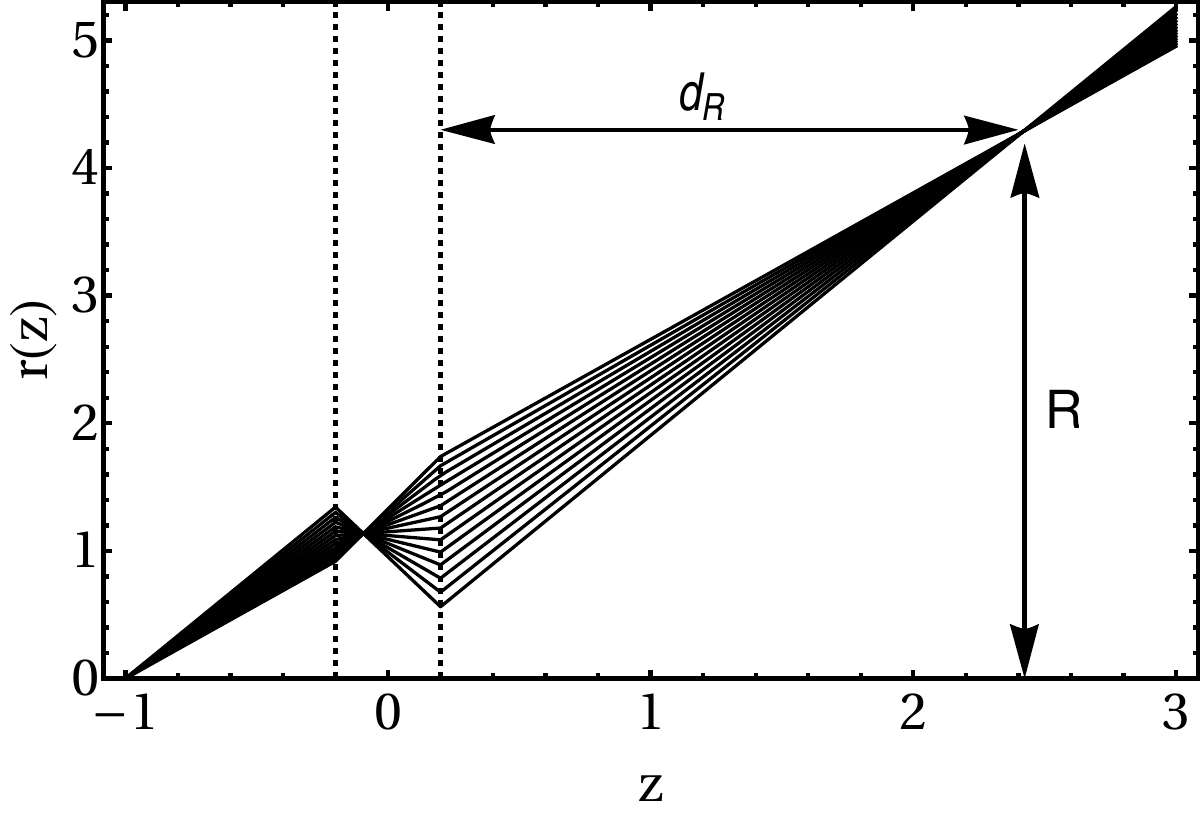}
		\caption{Example of electron focusing on a ring of radius $R$, at distance $d$: N-pNLS-N junction with a slab of thickness $L=0.8$, delimited by the vertical lines, and a source of monochromatic electrons in N on the left. The parameters used are $\varepsilon=-0.1$, $D_0=D=1$, $V_0=-2$, $r_0=0.1$, $d_s=0.8$.\label{fig:focus}}
	\end{figure}

	It is interesting now to look at the junction from an electron optics perspective. Based on the considerations of Sec. \ref{sec:N-NLS}, we propose here that the change in the band dispersion at the interface between a metal and a nodal-line semimetal creates naturally a negative refraction index, which can be used to redirect part of the electronic beam.
	For a sufficiently extended slab, see discussion above, we can for simplicity neglect the contribution of the surface states and study the geometric paths of the propagating electrons only. We consider a point source at a distance $d_{s}$ from the left interface of a slab of width $L$ and study the path of an electron incident on the slab with angle $\theta_{n}$. We are interested in the dependence of the radial distance $r$ on the distance $d$ measured from the right interface. With the considerations of Sec. \ref{sec:N-NLS}, we write
	\begin{equation}
		r\left(d,\theta_{n}\right)=d_{s}\tan\theta_{n}+L\tan\theta_{s}+d\tan\theta_{n} \; . 
	\end{equation}
	In Fig. \ref{fig:focus}, we show an interesting effect taking place for a $p$-doped NLS.  
	As discussed in the previous section, a focusing effect takes place in the $p$NLS, which is also present when the electrons exit the interface on the metallic side. Following similar geometric considerations, the condition that two paths with incidence angles $\theta_{n}$ and $\theta_{n}'$ meet at distance $d_R$ identifies the focal ring and can be cast into the form
 \begin{align}
		0 & \equiv r\left(d_R,\theta_{n}\right)-r\left(d_R,\theta'_{n}\right)	\\ \nonumber
         &=	\left(d_{s}+L\chi\left(\varepsilon\right)+d_R\right)  \left(\tan\theta_{n}-\tan\theta_{n}'\right) \;.
	\end{align}
	It follows that all the transmitted rays, independently of the incidence angle, cross at distance
 \begin{equation}\label{eq:dring}
		d_R=L\left| \chi\left(\varepsilon\right)\right|-d_{s}
\end{equation}
	from the second interface. Thus, they focus on a ring of radius
 \begin{equation}\label{eq:Rring}
		R=r\left(d_R\right)=L\left| \chi\left( \varepsilon\right) \right| \tan\theta_{n}^{*} \;,
\end{equation}
	which is a direct consequence of the inversion of the semiclassical velocity originated by the nodal line. Therefore, the result in Eq.\eqref{eq:Rring} provides a way of imaging the nodal line in real space. {The fact that the trajectories exactly meet on a ring is a consequence of the linearization of the momentum around the nodal line, while inclusion of higher-order terms would spread the focus onto a larger area \cite{Cheianov2007}. Nevertheless, when the chemical potential is reasonably close to the nodal line, the range of transverse momenta for which transmission is nonvanishing is narrow and the linearization is a relatively good approximation.}
	
\begin{figure}
		\includegraphics[height=0.18\textheight]{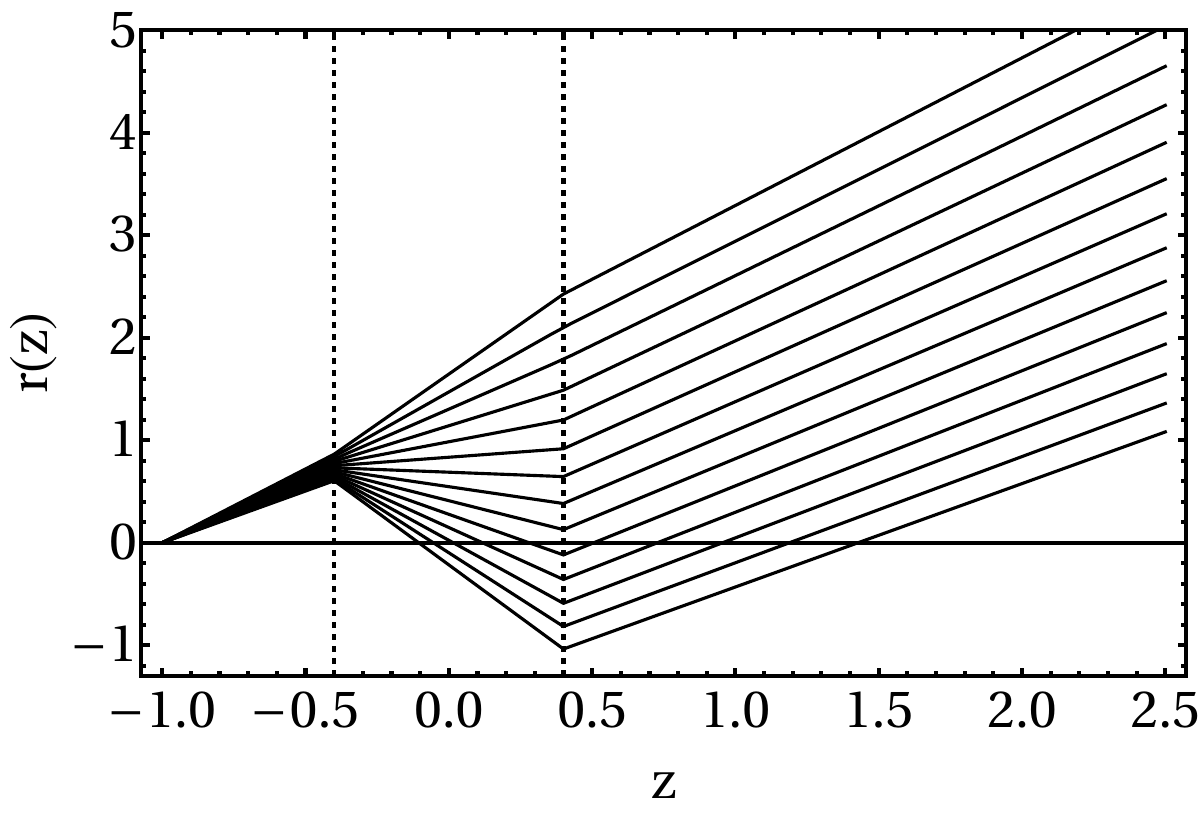}
		\caption{Observable effect of the negative refraction within the nodal line in a N-nNLS-N junction with a slab of thickness $L=0.8$ and a source of monochromatic electrons in N. In the N region on the right, a finite region of higher current density is present around the axis beyond the slab. The parameters used here are  $\varepsilon=0.1$, $D_0=D=1$, $V_0=-2$, $r_0=0.1$, $d_s=0.6$. \label{fig:conv} }
	\end{figure}

	In Fig. \ref{fig:conv}, we show instead a typical scenario occurring in a long junction including a $n$-doped NLS.
	As seen in Sec \ref{sec:N-NLS}, the refraction index for electrons with transverse momentum inside the nodal ring is negative, which implies that the electrons with tranverse momentum $q_\mathrm{p}<1$ are refracted back toward the axis while they travel within the NLS slab. After exiting in the metallic region again, they propagate with their original velocity, i.e., with the same angle with respect to the interface. However, traveling through the slab has shifted the path to a parallel one, which is evident in the spreading in Fig. \ref{fig:conv}. This suggests that a sequence of NLS slabs is an effective mean to keep electronic wavepackets close to the axis.
	
	\begin{figure}
		\includegraphics[height=0.18\textheight]{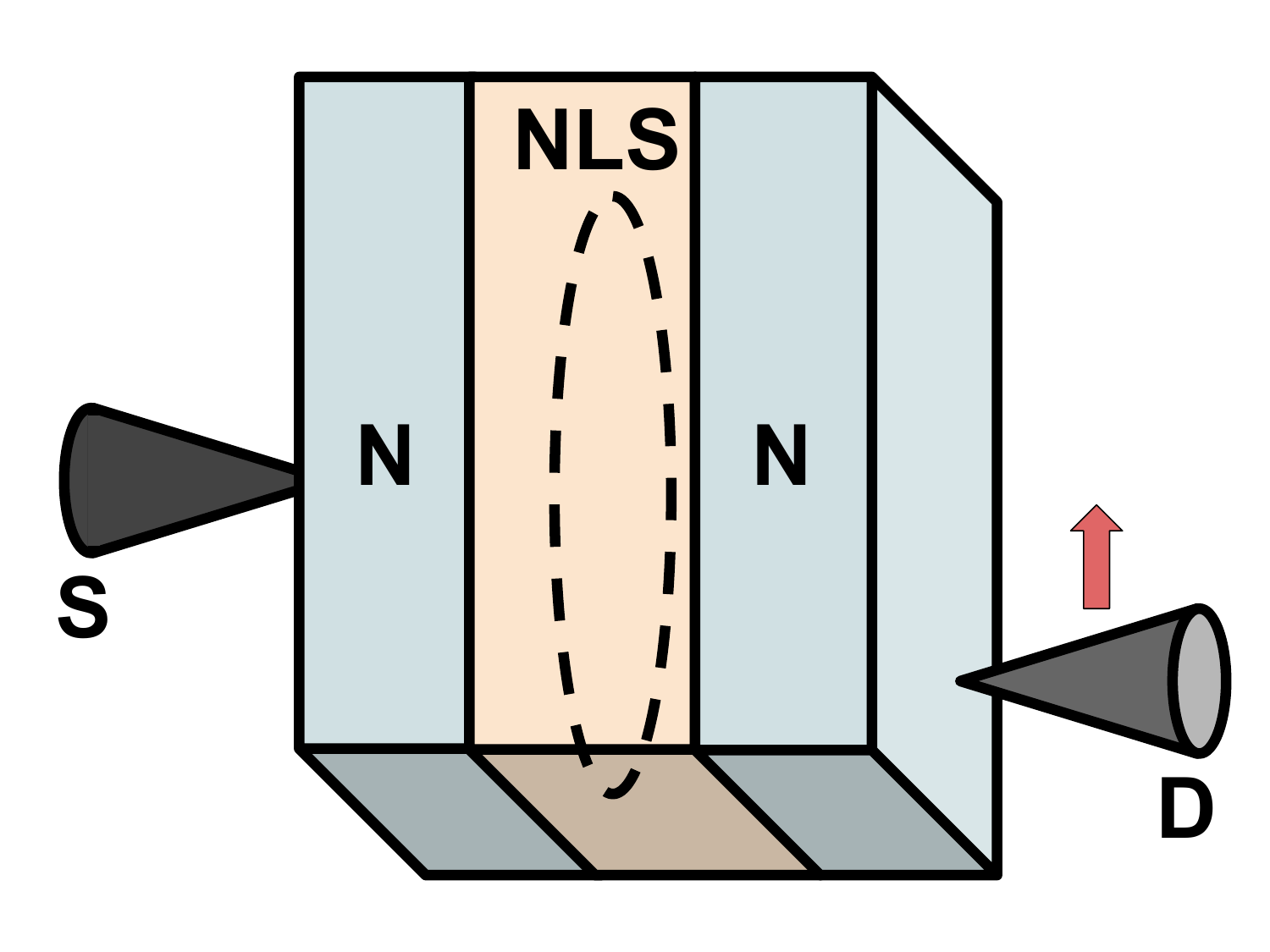}
		\caption{
			Proposed setup for the detection of the spatial current profile: an STM tip or a quantum point contact on one surface of the metallic sample (S) acts as a point source of electrons. After passing through the NLS slab, the current acquires an intensity distribution as a function of the distance from the axis, which can be probed by a mobile STM tip (D) on the other surface. \label{fig:STMdouble}
		}
	\end{figure}
	In order to probe these predictions, we propose the setup illustrated in Fig. \ref{fig:STMdouble}, in which the electric current through the N-NLS-N heterostructure is probed by means of two STM tips. In particular, one tip on one surface of the metallic sample acts as the localized source of electrons, while a mobile tip on the opposite surface can be used to scan for the local current density. 
	In the example of $\mbox{Ca}_3\mbox{P}_2$, with the parameters quoted in Sec.\ref{sec:slab} we obtain a reference energy scale $\hbar v k_\mathrm{NL} \approx 0.51 \mbox{eV}$ and taking $E\approx 0.095\mbox{ eV}$ \cite{Xie2015}, $E_0\approx 1\mbox{ eV}$, we obtain $\theta_n^*\approx 14.3$ degrees and $\chi\approx 5.1$.
	For a qualitative estimate, using instead $E = -0.1\mbox{ eV}$, $d_s = \mathcal{L} = 0.1$ $\mu\mbox{m}$ we obtain a focus on a ring of radius $R/k_{\mathrm{NL}} \approx 0.13$ $\mu\mbox{m}$ at distance $d_R/k_{\mathrm{NL}}\approx 0.41$ $\mu\mbox{m}$.
	
	\section{Conclusions}\label{sec:conclusions}
	We have studied in detail the effect of interfaces of nodal line semimetals in electronic transport, both for a metal-nodal line semimetal junction and for a nodal line semimetal slab embedded between two metallic samples.
	While the bulk topological invariant, quantized by a mirror symmetry, implies the existence of drumhead states on the surfaces parallel to the plane of the nodal line, 
    their support and penetration depth are also determined by the details of the termination.
	Such details determine the surface band dispersion, shifting it away from zero energy, and are of relevance in ARPES experiments targeting the drumhead states and for transport experiments, both along and across the surface.
	As transport in specific geometries of Weyl semimetals is strongly sensitive to boundary conditions \cite{DeMartino2021}, it will be interesting to check if the same holds true for nodal-line semimetals.
    	
	In the transparent limit, the interface becomes featureless, but the contribution of the surface states is still present in the slab geometry. In fact, we have shown that drumhead states hybridize into "drum" states and contribute to transport across the sample with a factor that scales like the inverse of the slab width. This extends the support of the transmission function, enhancing the conductivity across the slab and turning it finite even when the chemical potential is exactly at the band crossing. Interestingly, we have identified a series of resonances and connected them to the incidence angles of the electrons on the interface.
	
    We have shown that two kinds of refractions take place at the interfaces between a metal and a nodal-line semimetal, depending on the level of doping of the latter. In one regime, part of the electrons are refracted back toward the axis of the system; in the other, the electron paths cross on a ring of given radius, which can be exploited to image the nodal line in real space.
	Hole pockets in a NLS were detected, e.g., in \mbox{HfSiS} \cite{vanDelft2018}, but a more elaborate model is needed to describe the complicated Fermi surface of this material. We are presently not aware of experiments reporting a hole Fermi surface around the nodal line, but a material with such characteristic would open the possibility of exploring part of the physics described in this work.

    \buc{As a final remark, we only considered throughout this paper interfaces parallel to the plane of the nodal line. As nodal lines are protected by a crystal mirror symmetry, it is often the case that a material can be cleaved in such configuration \cite{Xie2015,Neupane2016,Liu2018x}. For a generic orientation, what matters is the projection of the nodal line onto the surface Brillouin zone, which delimits the surface states. The refraction effects only depend on the sign change of the group velocity  across the nodal line projection and therefore are expected to be qualitatively unaffected by the orientation. We do expect some changes in the formulas for the focus distance and radius, respectively Eqs. \eqref{eq:dring} and \eqref{eq:Rring}.}

	\begin{acknowledgments}
		We thank E. Zsurka and R. Egger for valuable advice, comments and feedback.
		M.R. thanks the Institut f\"ur Theoretische Physik IV of the HHU Düsseldorf and the Forschungszentrum Jülich for the hospitality, and acknowledges the Erasmus project and funding from INFN, gruppo collegato di Cosenza. F.B. acknowledges funding by the Deutsche Forschungsgemeinschaft under Projektnummer 277101999 - TRR 183 (project A02). K.M. acknowledges funding by the Bavarian Ministry of Economic Affairs, Regional Development and Energy within Bavaria’s High-Tech Agenda Project "Bausteine für das Quantencomputing auf Basis topologischer Materialien mit experimentellen und theoretischen Ansätzen" (grant no. 07 02/686 58/1/21 1/22 2/23).
	\end{acknowledgments}
	
	\appendix

	\section{Boundary Green's function}
	In the presence of a surface at $z=0$, the eigenstates in the region $z>0$ in the regime $m^2<\varepsilon^2$ have the form of reflected waves, with an incoming component $\sim e^{-iq_z z}$ and a reflected component $\sim e^{iq_z z}$. The reflection coefficient has unit modulus and is completely determined by the boundary condition in Eq. \eqref{eq:bc} as
	\begin{equation}\label{eq:r}
		r=e^{i\xi_{\alpha}\left(\varepsilon\right)}=-\frac{\left(\varepsilon-m\right)\cos\frac{\alpha}{2}+iq_{z}\sin\frac{\alpha}{2}}{\left(\varepsilon-m\right)\cos\frac{\alpha}{2}-iq_{z}\sin\frac{\alpha}{2}}
	\end{equation}
	(the label $\boldsymbol{q}_\mathrm{p}$ is omitted for compactness of notation). The Green's functions assume the generic form
	\begin{equation}\label{eq:GFbulk}
		G_{\varepsilon}\left(z,z';t\right)	= \sum_{r,r'=\pm}	G_{\varepsilon,r,r'}\left( t\right)  e^{iq_{z}\left(rz-r'z'\right)}e^{i(r-r')\xi_{\alpha}\left(\varepsilon\right)}\;,
	\end{equation}
	with the various components directly written from the bulk eigenstates \eqref{eq:eigenstates}. We are interested in computing this function on the boundary $z=z'=0$. In order to do this, we exploit the fixed spinor structure from the boundary condition \eqref{eq:bc} and write
	\begin{align}\label{eq:BGFb}
		G_{\varepsilon}^{R}\left(\boldsymbol{q}_{p};\omega\right)&
		=\frac{1}{\hbar v k_\mathrm{NL}L}\frac{\hat{g}_\alpha}{\omega+ i\eta-\varepsilon}\;,
	\end{align}
	in which $\eta$ denotes the small imaginary part of the frequency and the matrix $\hat{g}$ is
	\begin{equation}\label{eq:BGFmat}
		\hat{g}=\left(\begin{array}{cc}
			1+\cos\alpha & \sin\alpha\\
			\sin\alpha & 1-\cos\alpha
		\end{array}\right)\;.
	\end{equation}
	We have normalized the wavefunctions by a length $L$, which can be sent to infinity at the end.
	Similarly, one writes the boundary Green's function for the surface states as
		\begin{equation}\label{eq:BGFs}
			G_{s}^{R}\left(\boldsymbol{q}_\mathrm{p},z,z';\omega\right)=
			\frac{-1}{\hbar v k_\mathrm{NL}}\frac{\hat{g}_{\alpha} m_{\alpha}\left(q_{p}\right)\sin\alpha e^{ m_{\alpha}\left(q_{p}\right)\left(z+z'\right)\sin\alpha}}{\omega+ i\eta-m_{\alpha}\left(q_{p}\right)\cos\alpha}
		\end{equation}
	having taken advantage of the inverse penetration length \eqref{eq:kappainf} and of Eq. \eqref{eq:ssdispersion} for the energy  of the drumhead states.
	The associated local density of states per unit surface is then computed as
	\begin{equation}
		A\left(z,\omega\right) = -\frac{k_\mathrm{NL}^2}{\pi} \operatorname{Im}\left\lbrace\intop \frac{d^{2}\boldsymbol{q}_{p}}{\left(2\pi\right)^{2}} 	\mbox{Tr}\left[G^R_s\left(\boldsymbol{q}_{p},z,z;t\right)\right] \right\rbrace
	\end{equation}
	Substituting Eqs. \eqref{eq:BGFmat} and \eqref{eq:BGFs}, one obtains the expression \eqref{eq:surfaceLDoS} in the main text.
	
	\section{Quantization in a slab}\label{app:quantization}
	We provide here some detail about the quantization in a slab of Sec. \ref{sec:slab}. We apply the NLS Hamiltonian \eqref{eq:H} to the Ansatz
	\begin{equation}
		\psi_{q_{p}}\left(z\right)=g_{+}\left(z\right)\xi_{\alpha+}+g_{-}\left(z\right)\xi_{\alpha-},
	\end{equation}
	where $\xi_{\alpha\pm}$ denote the eigenvectors of $B\left(\alpha\right)$ with eigenvalue $\pm1$ and $g_\pm$ are two unknown functions of the coordinate across the slab. One obtains the decoupled equations
	\begin{equation}\label{eq:dequant}
		\left[m^{2}\left(q_{p}\right)-\varepsilon^{2}\right]g_{\pm}-\partial_{z}^{2}g_{\pm}=0
	\end{equation}
	for the unknown functions $g_\pm$, which have two families of solutions. If $\varepsilon^{2}>m^{2}\left(q_{p}\right)$, we have plane-wave solutions in the form $g_{\pm}=a_{\pm}e^{iq_zz}+b_{\pm}e^{-iq_zz}$ and momentum \mbox{$q_z=\sqrt{\varepsilon^{2}-m^{2}}$}. Further application of Eq. \eqref{eq:H} fixes the ratios between the coefficients as
	\begin{eqnarray}
		\frac{a_+}{a_-}&=&\frac{m\sin\alpha+iq_z}{m\cos\alpha-\varepsilon} \;, \\
		\frac{b_+}{b_-}&=&\frac{m\sin\alpha-iq_z}{m\cos\alpha-\varepsilon} \;.
	\end{eqnarray}
	The following step is imposing the boundary conditions. The application of Eq. \eqref{eq:bc} at $z=-L/2$ implies $g\left(-\frac{L}{2}\right)=0$, while the boundary condition at $L/2$ is equivalently written as
	\begin{equation}
		0	=	\xi_{-\alpha,-}^{\dagger}\cdot\psi\left(\frac{L}{2}\right)\;.
	\end{equation}
	Using
	\begin{align}\label{eq:rotate}
		&\xi_{-\alpha,+}= \cos\alpha\xi_{\alpha,+}+\sin\alpha\xi_{\alpha,-} ;, \nonumber\\
		&\xi_{-\alpha,+}= -\sin\alpha\xi_{\alpha,+}+\cos\alpha\xi_{\alpha,-} \;,
	\end{align}
	after some manipulation, one arrives at the quantization condition \eqref{eq:slabquantization}.
	
	In order to solve the quantization equation numerically, it is useful to have a good starting guess for the root-finding routine. With the parametrization
	\begin{equation}
		q_z=|m|\sin\chi\;,\qquad \varepsilon=|m|\cos\chi\;,\qquad 0\le\chi<\pi
	\end{equation}
	one notices that the quantization equation can be cast in the form
	\begin{equation}
		\tan\left(|m|L \sin\chi \right) =\frac{\sin\chi\sin\alpha}{\cos\chi\cos\alpha-\mbox{sign}\left( m\right) } \;.
	\end{equation}
	In the limit $|m|L\gg 1$ the solution is readily written as
	\begin{equation}
		q_z\approx \frac{\pi n}{L}\;, \qquad n=0,1,\ldots\;,
	\end{equation}
	which holds for the lowest bands $|n|\ll L/\pi$ and \mbox{$0<\alpha<\pi$}. This can be denoted a "bulk" limit, as the details of the boundaries do not matter. 
    In the opposite limit $|m|L\ll 1$, instead, one finds
	\begin{equation}
		q_z\approx \frac{\pi}{L}\left( n-\frac{1}{2}\right)\;, \qquad n=1,2,\ldots\;.
	\end{equation}

	If $m^2\left(q_{p}\right)>\varepsilon^{2}$, the differential equation \eqref{eq:dequant} admits evanescent solutions in the form $g_{\pm}=a_{\pm}e^{\kappa z}+b_{\pm}e^{-\kappa z}$. The energy is quantized by the condition
	\begin{equation}\label{eq:slabquantloc}
		\tanh\left(\kappa L\right)=\frac{\kappa\sin\alpha}{\varepsilon\cos\alpha-m}\;,
	\end{equation}
	with $\kappa=\sqrt{m^2-\varepsilon^2}$, which is just the analytic continuation of Eq. \eqref{eq:slabquantization}.
 The explicit form of the drum states is
	\begin{widetext}
		\begin{align}\label{eq:drum}
			\psi_{q_\mathrm{p}} \left(z\right) &= \mathcal{N}\Bigg\{ \left(\kappa\cosh\left[\kappa\left(z+\frac{L}{2}\right)\right]
			+m\sin\alpha\sinh\left[\kappa\left(z+\frac{L}{2}\right)\right]\right)\xi_{\alpha+}\cos\alpha 
			\nonumber\\
			&\qquad+\left(\kappa\coth\left(\kappa L\right)+m\sin\alpha\right)\sinh\left[\kappa\left(z+\frac{L}{2}\right)\right]\xi_{\alpha-}\sin\alpha\Bigg\} \;,
		\end{align}
	\end{widetext}
	up to a normalization $\mathcal{N}$, in which $\kappa$ is determined from the quantization condition \eqref{eq:slabquantloc} and the spinors $\xi_{\alpha,\pm}$ have been defined in Sec. \ref{sec:obc}. Using the relations \eqref{eq:rotate}, 
	it is straightforward to check that Eq. \eqref{eq:drum} satisfies the boundary conditions \eqref{eq:bc} and \eqref{eq:otherBC} and has equal weight around both surfaces.

	\section{Single and double interface, transparent limit}
	\subsection{Single interface}\label{app:singletransparent}
	We solve the quantum-mechanical problem of transmission through an interface, with the injection and detection of charge carriers taking place asymptotically far from the interface. To this end, we write the incoming, reflected and transmitted waves as
	\begin{equation}
		\begin{cases}\label{eq:singlesystem}
			\psi_{\nu_0q_0,\boldsymbol{q}_{p}} e^{\nu_0iq_{0}z}
			+r\psi_{-\nu_0q_0,\boldsymbol{q}_{p}}e^{-\nu_0iq_{0}z}
			\\			t\psi_{\nu q_z,\boldsymbol{q}_{p}} e^{ \nu i q_{z}z} \;
		\end{cases},
	\end{equation}
	where the bulk solutions are written in Eq. \eqref{eq:eigenstates}, $q_z$ and $q_0$ are the absolute values of the longitudinal momenta, defined as function of the energy in Eq. \eqref{eq:qq0def} and $\nu$, $\nu_0$ are the particle/hole indexes in defined in Sec. \ref{sec:N-NLS}. In the described setting, the states which need to be taken into account are only the ones that can carry current along $z$, i.e., the localized states are not included.
	The currents in the $z$ direction as derived from the Hamiltonian \eqref{eq:H} have the form
	\begin{equation}\label{eq:current}
		j_z = ev\tau_y
	\end{equation}
	and expectation values
	\begin{equation}\label{eq:currentev}
		j_\mathrm{NLS} = \frac{\nu ev q_z}{\varepsilon}\;,\qquad j_{0} = \frac{\nu_0evq_0}{\varepsilon_0}
	\end{equation}
	in the two regions. Matching the wavefunctions at the interface, one obtains two equations for the two components and can solve in the unknown reflection and transmission coefficients:
	\begin{align}\label{eq:singlesolution}
		r=&\frac{\nu\nu_{0}q_{0}\left(\varepsilon-m\right)-q_{z}\left(\varepsilon_{0}-m_{0}\right)}{\nu\nu_{0}q_{0}\left(\varepsilon-m\right)+q_{z}\left(\varepsilon_{0}-m_{0}\right)} \;,
		\\
		t=&\nu\sqrt{\frac{\varepsilon\left(\varepsilon-m\right)}{\varepsilon_{0}\left(\varepsilon_{0}-m_{0}\right)}}\frac{2q_{0}\left(\varepsilon_{0}-m_{0}\right)}{q_{0}\left|\varepsilon-m\right|+q_{z}\left(\varepsilon_{0}-m_{0}\right)} \;.
	\end{align}
	We note in passing that, within our normalization of the states, reflection and transmission satisfy the current conservation relation
	\begin{equation}
		\left|r\right|^2+\left|\frac{j_\text{NLS}}{j_0}\right|\left|t\right|^2 = 1 \;.
	\end{equation}
	The transmission probability $\mathcal{T}=\left|\frac{j_\text{NLS}}{j_0}\right|\left|t\right|^2$ is provided in Eq. \eqref{eq:Tsingletransparent}.

    Finally, we provide the inequalities corresponding to the transmission regimes discussed in Sec \ref{subsec:intmodel}.  For $\varepsilon>0$, the boundaries of regime (i), in which no transmission is possible, are 
    \begin{equation}
                0<\varepsilon<D\min\left\{\frac{D_0(1+r_0)+V_0}{D+D_0},1\right\} \;,
    \end{equation}
    reported as Eq. \eqref{eq:erange>0} in the main text. 
    The regime (iii), in which the highest transmission is found, is instead delimited by
    \begin{equation}
	\begin{cases}
		D\max\left\{0,\frac{D_0(1+r_0)+V_0}{D_0-D}\right\}<\varepsilon<D\;, & D>D_0\\
        0 \le \varepsilon<D\min\left\{-\frac{D_0(1+r_0)+V_0}{D_0-D},1\right\}\;, & D<D_0\;.
	\end{cases}
 	\end{equation}
    Eq. \eqref{eq:V0<smthg} must hold if $D \le D_0$, while the stricter condition $V_0<-(D+D_0r_0)$ is found for $D>D_0$.

	\subsection{Transport through a NLS slab}\label{app:slabtransparent}
	The scattering problem through a double interface is a standard scenario in quantum mechanics. We write the wavefunction as
	\begin{equation}
		\begin{cases}
			\psi_{\nu_0q_0,q_{p}}e^{\nu_{0}iq_{0}z}+r\psi_{-\nu_0q_0,q_{p}}e^{-\nu_{0}iq_{0}z} & z<-\frac{L}{2}\\
			a_{+}\psi_{q_z,q_{p}}e^{iq_{z}z}+a_{-}\psi_{-q_z,q_{p}}e^{-iq_{z}z} & -\frac{L}{2}<z<\frac{L}{2}\\
			t\psi_{\nu_0q_0,q_{p}}e^{\nu_{0}iq_{z}z} & z>\frac{L}{2}
		\end{cases}.
	\end{equation}
	Imposing the continuity at both interfaces, we determine the complex coefficients $r$, $t$, $a_\pm$ and  arrive at the end result in Eq. \eqref{eq:Tslab} for $\mathcal{T}=|t|^2$. We note in passing that it can be cast into the form
	\begin{equation}
		\mathcal{T}
		=\frac{\mathcal{T}_1^2}{1-2\mathcal{R}_1\cos\left(\varphi_\varepsilon\right)+\mathcal{R}_1^2}\,,
	\end{equation}
	where $\mathcal{T}_1$ is the transmission probability \eqref{eq:Tsingletransparent} through each interface and  $\mathcal{R}_1=1-\mathcal{T}_1$ is the corresponding reflection probability, while the phases $\varphi_\varepsilon$ has been defined in Eq. \eqref{eq:phase}. This is expected from our assumption of coherent transport through the slab \cite{Datta1997}.
	
	All thermoelectric coefficents are obtained from inversion of the response matrix (see, e.g., \cite{Grosso,Benenti2017,Pekola2021}). We focus in this work on the electric conductance per unit surface
	\begin{equation}\label{eq:sigma}
		\sigma = e^2 K_0 \, ,
	\end{equation}
	where
	\begin{equation}
		K_0 = \frac{1}{2\pi\hbar}\intop d\varepsilon  \mathcal{T}\left(\varepsilon\right) \left(-\frac{df_0}{d\varepsilon}\right) \,.
	\end{equation} 
	
	In the expression above, $\mathcal{T}$ is the transmission function at given energy defined in Eq. \ref{eq:T(e)} and $f_0$ the free electronic distribution from Eq. \eqref{eq:f0}.		
	
	\section{Angles}\label{app:angles}
	In this appendix, we provide some details of the calculation leading to Eq. \eqref{eq:Snell}. Assuming a small Fermi surface around the nodal line, see Sec. \ref{sec:NLS}, we can consider momenta close to the nodal line $q_{p}\approx 1 $ and retain only the linear term in the deviation $\delta q_{p}$, i.e., $q_{p} = 1+\delta q_{p}$. Throughout this section, we will therefore write the velocities (in units of $v$) and all the expressions up to terms $\sim O\left(\delta q_{p}^{2}\right)$. We begin by writing the semiclassical velocity components \eqref{eq:us}  in the NLS to this order
	\begin{equation}
		u_{p}\approx\frac{4D^{2}}{\varepsilon}\delta q_{p} \,\qquad
		|u_{z}|\approx\,1 \;.
	\end{equation}
	Using the expressions above, we relate the exit angle in Eq. \eqref{eq:us} to the distance from the nodal line \begin{equation}\label{eq:tanthetas}
		\tan\theta_{s}\,=\,\frac{u_{p}}{u_{z}}	\approx	\frac{4D^2}{\varepsilon}\delta q_{p} \;.
	\end{equation}
	The strategy is now to express $\delta q_{p}$, conserved across the interface, 
    in terms of the energy and angle of the incoming particle. From the definition \eqref{eq:ws} we obtain
 \begin{align}
		\frac{w_{p}}{v}&\approx	\frac{2D_0^{2}}{\varepsilon_0 }\left(1+r_{0}\right)\left( 1+\frac{3+r_{0}}{1+r_{0}}\delta q_{p}\right) \;,
		\\
	\frac{w_{z}}{v}&\approx \frac{q_0(1)}{\varepsilon_0}
  \left(1-\frac{2D_{0}^{2}\left(1+r_{0}\right)}{q_0^2(1) 
  }\delta q_{p}\right) \;.
		\nonumber
\end{align}
	Note that the last line never changes sign, as long as $\left|\delta q_{p}\right|\ll1$. Combining these relations, we obtain
 \begin{equation}\label{eq:deltakp}
		\delta q_{p}=\frac{q_0^3(1)}{2D_{0}^{2}}\frac{
			\left(\tan\theta_{n}-\tan\theta_n^*\right)}{2\varepsilon_{0}^{2}+\left(1+r_{0}\right) q_0^2(1)},
	\end{equation}
	where $\theta_{n}^{*}$ is defined in Eq. \eqref{eq:thetastar}. We can now substitute this expression into Eq. \eqref{eq:tanthetas} and obtain the "generalized Snell's law" \eqref{eq:Snell} in the main text.
	
	Interestingly, the condition that the electron has a real momentum $q_z$ in the NLS identifies a range of transverse momenta \eqref{eq:qrange}, which can be promptly translated into a range of allowed incidence angles from the N side for which transmission into the NLS is possible. Substitution of Eq. \eqref{eq:deltakp} into Eq. \eqref{eq:qrange} implies that 
	\begin{equation}\label{eq:thetarange}
		\tan\theta_\mathrm{min} < \tan\theta_n < \tan\theta_\mathrm{max},
	\end{equation}
	with
 \begin{widetext}
\begin{align} \label{eq:tanthetaminmax}
			\tan\theta_\mathrm{min} & = \frac{2D_{0}^{2}}{q_0^3(1)} \left[ \left( 1+r_{0} \right) q_0^2(1) - 
   \frac{\left| \varepsilon \right|}{2D} \left( 2\varepsilon_0^2+(1+r_0)q_0^2(1) \right) \right]  \;, \\
			\tan\theta_\mathrm{max} &= \frac{2D_{0}^{2}}{q_0^3(1)}\left[ \left(1+r_{0}\right) q_0^2(1) + 
   \frac{\left| \varepsilon \right|}{2D} \left( 2\varepsilon_0^2+(1+r_0)q_0^2(1) \right) \right]  \;.
    	\end{align}
 \end{widetext}

	\bibliography{NLI.bib}
	
\end{document}